\documentclass[3p]{elsarticle}

\usepackage{amssymb}
\usepackage{amsmath}
\usepackage{subcaption}
\newcommand{\micron}{$\mathrm{\mu m}$}

\newcommand{\totalarea}{6.6}
\newcommand{\MIP}{proton}

\newcommand{\aligneventss}{15} %15.5
\newcommand{\hitsignal}{$S\times \mathrm{cos}(\theta)$}
\newcommand{\tempdegree}{$^\circ\mathrm{C}$}
\newcommand{\eventsforalignmentdaily}{\mbox{0.8$\mathrm{M}$}} % 0.79
\newcommand{\fixedchisqare}{\mbox{\emph{fixed}-$\chi^2$}}
\newcommand{\variablechisqare}{\mbox{\emph{variable}-$\chi^2$}}
\newcommand{\highzuncertainty}{Systematic uncertainty due to the contribution of nuclei with charge $Z>1$ in the alignment sample is less than 0.1~\micron~and 0.2~\micron~for~$\sigma_1$~and~$\sigma_{12}$~respectively, as described in Section~\ref{sec:syst}}
\newcommand{\errorbarsdescribtion}{Data error bars include statistical uncertainties and systematic uncertainties due to on-orbit variation of the alignment and multiple scattering, summed in quadrature. Monte-Carlo error bars are statistical uncertainties only, as described in Section~\ref{sec:syst}}

\makeatletter
\def\@fnsymbol#1{\ensuremath{\ifcase#1\or \dagger\or \ddagger\or
    \mathsection\or \mathparagraph\or \|\or **\or \dagger\dagger
    \or \ddagger\ddagger \else\@ctrerr\fi}}
\makeatletter

\usepackage{etoolbox}
\makeatletter
\patchcmd{\ps@pprintTitle}% <cmd>
  {Preprint submitted}% <search>
  {} %To be submitted}% <replace>
  {}{}% <succes><failure>
\patchcmd{\ps@pprintTitle}% <cmd>
  {Elsevier}% <search>
  {} %To be submitted}% <replace>
  {}{}% <succes><failure>
\patchcmd{\ps@pprintTitle}% <cmd>
  {to}% <search>
  {} %To be submitted}% <replace>
  {}{}% <succes><failure>
\makeatother

\begin{document}

%\linenumbers

\begin{frontmatter}

\title{Internal alignment and position resolution of the silicon tracker of DAMPE determined with orbit data}
%\title{Internal alignment and position resolution of the silicon tracker of DAMPE determined with the orbit data}

\author[gva]{A. Tykhonov\corref{cor1}}
\cortext[cor1]{Corresponding author}
\ead{andrii.tykhonov@cern.ch}

\author[infn_pga]{G. Ambrosi}
%\addressIstituto Nazionale di Fisica Nucleare Sezione di Perugia, I-06123 Perugia, Italy}

%\author[ustc]{Q. An}
%%\addressState Key Laboratory of Particle Detection and Electronics, University of Science and Technology of China, Hefei 230026, China}

\author[gva]{R. Asfandiyarov}
%%\addressDepartment of Nuclear and Particle Physics, University of Geneva, CH-1211, Switzerland}

\author[gva]{P. Azzarello}
%%\addressDepartment of Nuclear and Particle Physics, University of Geneva, CH-1211, Switzerland}

\author[mat_lec,infn_lec]{P. Bernardini}
%%\addressUniversit\`a del Salento - Dipartimento di Matematica e Fisica "E. De Giorgi", I-73100, Lecce, Italy}
%%\addressIstituto Nazionale di Fisica Nucleare (INFN) - Sezione di Lecce , I-73100 , Lecce, Italy}

\author[infn_pga,geo_pga]{B. Bertucci}
%%\addressDipartimento di Fisica e Geologia, Universit\`a degli Studi di Perugia, I-06123 Perugia, Italy}
%%\addressIstituto Nazionale di Fisica Nucleare Sezione di Perugia, I-06123 Perugia, Italy}

\author[infn_pga,geo_pga]{A. Bolognini\footnote{Now at RUAG Space, Zurich, Switzerland}} %Stauffacherstrasse 65, 3000 Bern 22 Switzerland}}

\author[gva]{F. Cadoux}

%\author[pmo]{M. S. Cai}
%%\addressKey Laboratory of Dark Matter and Space Astronomy, Purple Mountain Observatory, Chinese Academy of Sciences, Nanjing 210008, China}

%\author[infn_bari]{M. Caragiulo}
%%\addressIstituto Nazionale di Fisica Nucleare Sezione di Bari, I-70125, Bari, Italy}

%\author[pmo]{J. Chang}
%\addressKey Laboratory of Dark Matter and Space Astronomy, Purple Mountain Observatory, Chinese Academy of Sciences, Nanjing 210008, China}

%\author[pmo,ucas]{D. Y. Chen}
%\addressKey Laboratory of Dark Matter and Space Astronomy, Purple Mountain Observatory, Chinese Academy of Sciences, Nanjing 210008, China}
%\addressUniversity of Chinese Academy of Sciences, Yuquan Road 19, Beijing 100049, China}

%\author[ustc]{H. F. Chen}
%\addressState Key Laboratory of Particle Detection and Electronics, University of Science and Technology of China, Hefei 230026, China}

%\author[imp]{J. L. Chen}
%\addressInstitute of Modern Physics, Chinese Academy of Sciences, Nanchang Road 59, Lanzhou 730000, China}

%\author[pmo,ucas]{W. Chen}
%\addressKey Laboratory of Dark Matter and Space Astronomy, Purple Mountain Observatory, Chinese Academy of Sciences, Nanjing 210008, China}
%\addressUniversity of Chinese Academy of Sciences, Yuquan Road 19, Beijing 100049, China}

%\author[pmo]{M. Y. Cui}
%\addressKey Laboratory of Dark Matter and Space Astronomy, Purple Mountain Observatory, Chinese Academy of Sciences, Nanjing 210008, China}

%\author[space_beij]{T. S. Cui}
%\addressNational Space Science Center, Chinese Academy of Sciences, Nanertiao 1, Zhongguancun, Haidian district, Beijing 100190, China}

\author[mat_lec,infn_lec]{A.~D'Amone}
%\addressUniversit\`a del Salento - Dipartimento di Matematica e Fisica "E. De Giorgi", I-73100, Lecce, Italy}
%\addressIstituto Nazionale di Fisica Nucleare (INFN) - Sezione di Lecce , I-73100 , Lecce, Italy}

\author[mat_lec,infn_lec]{A.~De~Benedittis}
%\addressUniversit\`a del Salento - Dipartimento di Matematica e Fisica "E. De Giorgi", I-73100, Lecce, Italy}
%\addressIstituto Nazionale di Fisica Nucleare (INFN) - Sezione di Lecce , I-73100 , Lecce, Italy}

\author[infn_lec,gran_sasso]{I.~De~Mitri}
%\author[mat_lec,infn_lec]{I.~De~Mitri}
%\addressUniversit\`a del Salento - Dipartimento di Matematica e Fisica "E. De Giorgi", I-73100, Lecce, Italy}
%\addressIstituto Nazionale di Fisica Nucleare (INFN) - Sezione di Lecce , I-73100 , Lecce, Italy}

\author[mat_lec,infn_lec]{M.~Di~Santo}

%\author[ustc]{J. N. Dong}
%\addressState Key Laboratory of Particle Detection and Electronics, University of Science and Technology of China, Hefei 230026, China}

%\author[pmo]{T. K. Dong}
%\addressKey Laboratory of Dark Matter and Space Astronomy, Purple Mountain Observatory, Chinese Academy of Sciences, Nanjing 210008, China}

\author[ihep]{Y. F. Dong}
%\addressInstitute of High Energy Physics, Chinese Academy of Sciences, YuquanLu 19B, Beijing 100049,  China}

%\author[space_beij]{Z. X. Dong}
%\addressNational Space Science Center, Chinese Academy of Sciences, Nanertiao 1, Zhongguancun, Haidian district, Beijing 100190, China}

%\author[pmo,ucas]{K. K. Duan}
%\addressKey Laboratory of Dark Matter and Space Astronomy, Purple Mountain Observatory, Chinese Academy of Sciences, Nanjing 210008, China}
%\addressUniversity of Chinese Academy of Sciences, Yuquan Road 19, Beijing 100049, China}

%\author[imp]{J. L. Duan}
%\addressInstitute of Modern Physics, Chinese Academy of Sciences, Nanchang Road 59, Lanzhou 730000, China}

\author[infn_pga]{M. Duranti}
%\addressDipartimento di Fisica e Geologia, Universit\`a degli Studi di Perugia, I-06123 Perugia, Italy}
%\addressIstituto Nazionale di Fisica Nucleare Sezione di Perugia, I-06123 Perugia, Italy}

%\author[infn_pga,asi]{D. D'Urso}
\author[sassari,catania,asi]{D. D'Urso}
%\addressIstituto Nazionale di Fisica Nucleare Sezione di Perugia, I-06123 Perugia, Italy}
%\addressASI Science Data Center (ASDC), I-00133 Roma, Italy}

\author[ihep]{R. R. Fan}
%\addressInstitute of High Energy Physics, Chinese Academy of Sciences, YuquanLu 19B, Beijing 100049,  China}

%\author[pmo]{Y. Z. Fan}
%\addressKey Laboratory of Dark Matter and Space Astronomy, Purple Mountain Observatory, Chinese Academy of Sciences, Nanjing 210008, China}

%\author[imp]{F. Fang}
%\addressInstitute of Modern Physics, Chinese Academy of Sciences, Nanchang Road 59, Lanzhou 730000, China}

%\author[ustc]{C. Q. Feng}
%\addressState Key Laboratory of Particle Detection and Electronics, University of Science and Technology of China, Hefei 230026, China}

%\author[pmo]{L. Feng}
%\addressKey Laboratory of Dark Matter and Space Astronomy, Purple Mountain Observatory, Chinese Academy of Sciences, Nanjing 210008, China}

\author[infn_bari,fis_bari]{P. Fusco}
%\addressIstituto Nazionale di Fisica Nucleare Sezione di Bari, I-70125, Bari, Italy}
%\addressDipartimento di Fisica "M.Merlin" dell'Univerisity e del Politecnico di Bari, I-70126, Bari, Italy}

\author[gva]{V. Gallo}
%\addressDepartment of Nuclear and Particle Physics, University of Geneva, CH-1211, Switzerland}

%\author[ustc]{F. J. Gan}
%\addressState Key Laboratory of Particle Detection and Electronics, University of Science and Technology of China, Hefei 230026, China}

\author[ihep]{M. Gao}
%\addressInstitute of High Energy Physics, Chinese Academy of Sciences, YuquanLu 19B, Beijing 100049,  China}

%\author[ustc]{S. S. Gao}
%\addressState Key Laboratory of Particle Detection and Electronics, University of Science and Technology of China, Hefei 230026, China}

\author[infn_bari]{F. Gargano}
%\addressIstituto Nazionale di Fisica Nucleare Sezione di Bari, I-70125, Bari, Italy}

\author[infn_pga,geo_pga]{S. Garrappa}

\author[ihep]{K. Gong}
%\addressInstitute of High Energy Physics, Chinese Academy of Sciences, YuquanLu 19B, Beijing 100049,  China}

%\author[pmo]{Y. Z. Gong}
%\addressKey Laboratory of Dark Matter and Space Astronomy, Purple Mountain Observatory, Chinese Academy of Sciences, Nanjing 210008, China}

%\author[fis_pga,infn_pga]{M. Graziani}
%\addressDipartimento di Fisica e Geologia, Universit\`a degli Studi di Perugia, I-06123 Perugia, Italy}
%\addressIstituto Nazionale di Fisica Nucleare Sezione di Perugia, I-06123 Perugia, Italy}

%\author[pmo]{J. H. Guo}
%\footnote{The corresponding author (jhguo@pmo.ac.cn).}}
%\addressKey Laboratory of Dark Matter and Space Astronomy, Purple Mountain Observatory, Chinese Academy of Sciences, Nanjing 210008, China}

%\author[pmo,ucas]{Y. M. Hu}
%\addressKey Laboratory of Dark Matter and Space Astronomy, Purple Mountain Observatory, Chinese Academy of Sciences, Nanjing 210008, China}
%\addressUniversity of Chinese Academy of Sciences, Yuquan Road 19, Beijing 100049, China}

%\author[ustc]{G. S. Huang}
%\addressState Key Laboratory of Particle Detection and Electronics, University of Science and Technology of China, Hefei 230026, China}

%\author[pmo]{Y. Y. Huang}
%\addressKey Laboratory of Dark Matter and Space Astronomy, Purple Mountain Observatory, Chinese Academy of Sciences, Nanjing 210008, China}

\author[infn_pga]{M. Ionica}
%\addressIstituto Nazionale di Fisica Nucleare Sezione di Perugia, I-06123 Perugia, Italy}

%\author[ustc]{D. Jiang}
%\addressState Key Laboratory of Particle Detection and Electronics, University of Science and Technology of China, Hefei 230026, China}

%\author[pmo,ucas]{W. Jiang}
%\addressKey Laboratory of Dark Matter and Space Astronomy, Purple Mountain Observatory, Chinese Academy of Sciences, Nanjing 210008, China}
%\addressUniversity of Chinese Academy of Sciences, Yuquan Road 19, Beijing 100049, China}

%\author[ustc]{X. Jin}
%\addressState Key Laboratory of Particle Detection and Electronics, University of Science and Technology of China, Hefei 230026, China}

%\author[imp]{J. Kong}
%\addressInstitute of Modern Physics, Chinese Academy of Sciences, Nanchang Road 59, Lanzhou 730000, China}

\author[gva]{D. La Marra}

\author[pmo]{S. J. Lei}
%\addressKey Laboratory of Dark Matter and Space Astronomy, Purple Mountain Observatory, Chinese Academy of Sciences, Nanjing 210008, China}

%\author[pmo,ucas]{S. Li}
%\addressKey Laboratory of Dark Matter and Space Astronomy, Purple Mountain Observatory, Chinese Academy of Sciences, Nanjing 210008, China}
%\addressUniversity of Chinese Academy of Sciences, Yuquan Road 19, Beijing 100049, China}

\author[pmo]{X. Li}

\author[infn_bari,fis_bari]{F. Loparco}
%\addressIstituto Nazionale di Fisica Nucleare Sezione di Bari, I-70125, Bari, Italy}
%\addressDipartimento di Fisica "M.Merlin" dell'Univerisity e del Politecnico di Bari, I-70126, Bari, Italy}

%\author[space_beij]{M. Ma}
%\addressNational Space Science Center, Chinese Academy of Sciences, Nanertiao 1, Zhongguancun, Haidian district, Beijing 100190, China}

%\author[pmo]{P. X. Ma}
%\addressKey Laboratory of Dark Matter and Space Astronomy, Purple Mountain Observatory, Chinese Academy of Sciences, Nanjing 210008, China}

%\author[ustc]{S. Y. Ma}
%\addressState Key Laboratory of Particle Detection and Electronics, University of Science and Technology of China, Hefei 230026, China}

%\author[pmo]{T. Ma}
%\addressKey Laboratory of Dark Matter and Space Astronomy, Purple Mountain Observatory, Chinese Academy of Sciences, Nanjing 210008, China}

%\author[space_beij]{X. Q. Ma}
%\addressNational Space Science Center, Chinese Academy of Sciences, Nanertiao 1, Zhongguancun, Haidian district, Beijing 100190, China}

%\author[space_beij]{X. Y. Ma}
%\addressNational Space Science Center, Chinese Academy of Sciences, Nanertiao 1, Zhongguancun, Haidian district, Beijing 100190, China}

\author[mat_lec,infn_lec]{G. Marsella}
%\addressUniversit\`a del Salento - Dipartimento di Matematica e Fisica "E. De Giorgi", I-73100, Lecce, Italy}
%\addressIstituto Nazionale di Fisica Nucleare (INFN) - Sezione di Lecce , I-73100 , Lecce, Italy}

\author[infn_bari]{M.N. Mazziotta}
%\addressIstituto Nazionale di Fisica Nucleare Sezione di Bari, I-70125, Bari, Italy}

%\author[imp]{D. Mo}
%\addressInstitute of Modern Physics, Chinese Academy of Sciences, Nanchang Road 59, Lanzhou 730000, China}

%\author[imp]{X. Y. Niu}
%\addressInstitute of Modern Physics, Chinese Academy of Sciences, Nanchang Road 59, Lanzhou 730000, China}

%\author[pmo]{X. Y. Peng}
%\addressKey Laboratory of Dark Matter and Space Astronomy, Purple Mountain Observatory, Chinese Academy of Sciences, Nanjing 210008, China}

\author[ihep]{W. X. Peng}
%\addressInstitute of High Energy Physics, Chinese Academy of Sciences, YuquanLu 19B, Beijing 100049,  China}

\author[ihep]{R. Qiao}
%\addressInstitute of High Energy Physics, Chinese Academy of Sciences, YuquanLu 19B, Beijing 100049,  China}

%\author[space_beij]{J. N. Rao}
%\addressNational Space Science Center, Chinese Academy of Sciences, Nanertiao 1, Zhongguancun, Haidian district, Beijing 100190, China}

\author[gva]{M. M. Salinas}
%\addressDepartment of Nuclear and Particle Physics, University of Geneva, CH-1211, Switzerland}

%\author[space_beij]{G. Z. Shang}
%\addressNational Space Science Center, Chinese Academy of Sciences, Nanertiao 1, Zhongguancun, Haidian district, Beijing 100190, China}

%\author[space_beij]{W. H. Shen}
%\addressNational Space Science Center, Chinese Academy of Sciences, Nanertiao 1, Zhongguancun, Haidian district, Beijing 100190, China}

%\author[pmo,ucas]{Z. Q. Shen}
%\addressKey Laboratory of Dark Matter and Space Astronomy, Purple Mountain Observatory, Chinese Academy of Sciences, Nanjing 210008, China}
%\addressUniversity of Chinese Academy of Sciences, Yuquan Road 19, Beijing 100049, China}

%\author[ustc]{Z. T. Shen}
%\addressState Key Laboratory of Particle Detection and Electronics, University of Science and Technology of China, Hefei 230026, China}

%\author[space_beij]{J. X. Song}
%\addressNational Space Science Center, Chinese Academy of Sciences, Nanertiao 1, Zhongguancun, Haidian district, Beijing 100190, China}

%\author[imp]{H. Su}
%\addressInstitute of Modern Physics, Chinese Academy of Sciences, Nanchang Road 59, Lanzhou 730000, China}

%\author[pmo,hk]{M. Su}
%\addressDepartment of Physics, The University of Hongkong, Pokfulam Road, Hong Kong}}
%\addressKey Laboratory of Dark Matter and Space Astronomy, Purple Mountain Observatory, Chinese Academy of Sciences, Nanjing 210008, China}

%\author[imp]{Z. Y. Sun}
%\addressInstitute of Modern Physics, Chinese Academy of Sciences, Nanchang Road 59, Lanzhou 730000, China}

\author[infn_lec]{A. Surdo}
%\addressIstituto Nazionale di Fisica Nucleare (INFN) - Sezione di Lecce, I-73100, Lecce, Italy}

%\author[space_beij]{X. J. Teng}
%\addressNational Space Science Center, Chinese Academy of Sciences, Nanertiao 1, Zhongguancun, Haidian district, Beijing 100190, China}

%\author[space_beij]{X. B. Tian}
%\addressNational Space Science Center, Chinese Academy of Sciences, Nanertiao 1, Zhongguancun, Haidian district, Beijing 100190, China}

%\author[gva]{A. Tykhonov\corref{cor1}}
%\ead{andrii.tykhonov@cern.ch}

\author[infn_pga,geo_pga]{V. Vagelli}

\author[gva]{S. Vitillo}
%\addressDepartment of Nuclear and Particle Physics, University of Geneva, CH-1211, Switzerland}

%\author[ustc]{C. Wang}
%\addressState Key Laboratory of Particle Detection and Electronics, University of Science and Technology of China, Hefei 230026, China}

%\author[space_beij]{H. Wang}
%\addressNational Space Science Center, Chinese Academy of Sciences, Nanertiao 1, Zhongguancun, Haidian district, Beijing 100190, China}

\author[ihep]{H. Y. Wang}
%\addressInstitute of High Energy Physics, Chinese Academy of Sciences, YuquanLu 19B, Beijing 100049,  China}

\author[ihep]{J. Z. Wang}
%\addressInstitute of High Energy Physics, Chinese Academy of Sciences, YuquanLu 19B, Beijing 100049,  China}

\author[infn_lec,gran_sasso]{Z~.M.~Wang}

\author[ihep]{D. Wu}
%\addressInstitute of High Energy Physics, Chinese Academy of Sciences, YuquanLu 19B, Beijing 100049,  China}

%\author[pmo]{J. Wu}
%\addressKey Laboratory of Dark Matter and Space Astronomy, Purple Mountain Observatory, Chinese Academy of Sciences, Nanjing 210008, China}

%\author[ustc]{L. B. Wu}
%\addressState Key Laboratory of Particle Detection and Electronics, University of Science and Technology of China, Hefei 230026, China}

%\author[space_beij]{S. S. Wu}
%\addressNational Space Science Center, Chinese Academy of Sciences, Nanertiao 1, Zhongguancun, Haidian district, Beijing 100190, China}

\author[gva]{X. Wu}

\author[ihep]{F. Zhang}
%\addressInstitute of High Energy Physics, Chinese Academy of Sciences, YuquanLu 19B, Beijing 100049,  China}

%\author[ustc]{J. B. Zhang}
%\addressState Key Laboratory of Particle Detection and Electronics, University of Science and Technology of China, Hefei 230026, China}

\author[ihep]{J. Y. Zhang}

\author[ihep]{H. Zhao}
%\addressInstitute of High Energy Physics, Chinese Academy of Sciences, YuquanLu 19B, Beijing 100049, China}

%\author[space_beij]{X. F. Zhao}
%\addressNational Space Science Center, Chinese Academy of Sciences, Nanertiao 1, Zhongguancun, Haidian district, Beijing 100190, China}

%\author[space_beij]{C. Y. Zhou}
%\addressNational Space Science Center, Chinese Academy of Sciences, Nanertiao 1, Zhongguancun, Haidian district, Beijing 100190, China}

%\author[imp]{Y. Zhou}
%\addressInstitute of Modern Physics, Chinese Academy of Sciences, Nanchang Road 59, Lanzhou 730000, China}

%\author[ustc]{X. Zhu}
%\addressState Key Laboratory of Particle Detection and Electronics, University of Science and Technology of China, Hefei 230026, China}

%\author[space_beij]{Y. Zhu}
%\addressNational Space Science Center, Chinese Academy of Sciences, Nanertiao 1, Zhongguancun, Haidian district, Beijing 100190, China}

\author[gva]{S. Zimmer\footnote{Now at University of Innsbruck, Austria}}
%\addressDepartment of Nuclear and Particle Physics, University of Geneva, CH-1211, Switzerland}

\address[gva]{Department of Nuclear and Particle Physics, University of Geneva, CH-1211, Switzerland}
%\address[pmo]{Key Laboratory of Dark Matter and Space Astronomy, Purple Mountain Observatory, Chinese Academy of Sciences, Nanjing 210008, China}
\address[infn_pga]{Istituto Nazionale di Fisica Nucleare Sezione di Perugia, I-06123 Perugia, Italy}
%\address[ustc]{State Key Laboratory of Particle Detection and Electronics, University of Science and Technology of China, Hefei 230026, China}
%\address{Department of Nuclear and Particle Physics, University of Geneva, CH-1211, Switzerland}
\address[mat_lec]{Universit\`a del Salento - Dipartimento di Matematica e Fisica "E. De Giorgi", I-73100, Lecce, Italy}
\address[infn_lec]{Istituto Nazionale di Fisica Nucleare (INFN) -- Sezione di Lecce , I-73100 , Lecce, Italy}
\address[geo_pga]{Dipartimento di Fisica e Geologia, Universit\`a degli Studi di Perugia, I-06123 Perugia, Italy}
\address[gran_sasso]{Gran Sasso Science Institute (GSSI), Via Iacobucci 2, I-67100, L'Aquila, Italy}
%\address[ucas]{University of Chinese Academy of Sciences, Yuquan Road 19, Beijing 100049, China}
%\address[imp]{Institute of Modern Physics, Chinese Academy of Sciences, Nanchang Road 59, Lanzhou 730000, China}
%\address[space_beij]{National Space Science Center, Chinese Academy of Sciences, Nanertiao 1, Zhongguancun, Haidian district, Beijing 100190, China}
\address[ihep]{Institute of High Energy Physics, Chinese Academy of Sciences, YuquanLu 19B, Beijing 100049, China}
\address[sassari]{Chemistry and Pharmacy Department, Universit\`a degli Studi di Sassari, Sassari 07100, Italy}
\address[catania]{Istituto Nazionale di Fisica Nucleare (INFN) --  Laboratori Nazionali del Sud, Catania 95123, Italy}
%\address[asi]{ASI Science Data Center (ASDC), I-00133 Roma, Italy}
\address[asi]{SSDC-ASI via del Politecnico snc, 00133 Roma, Italy}
\address[infn_bari]{Istituto Nazionale di Fisica Nucleare Sezione di Bari, I-70125, Bari, Italy}
\address[fis_bari]{Dipartimento di Fisica "M.Merlin" dell'Univerisit\`a e del Politecnico di Bari, I-70126, Bari, Italy}
\address[pmo]{Key Laboratory of Dark Matter and Space Astronomy, Purple Mountain Observatory, Chinese Academy of Sciences, Nanjing 210008, China}

\begin{abstract}
The DArk Matter Particle Explorer (DAMPE)  is a 
 space-borne particle detector designed to probe electrons and gamma-rays in the few GeV to \mbox{10 TeV} energy range, as well as cosmic-ray proton and nuclei components between \mbox{10 GeV} and \mbox{100 TeV}. The silicon-tungsten tracker-converter is a crucial component of DAMPE. 
 It allows the direction of incoming photons converting into electron-positron pairs to be estimated, and the trajectory and charge (Z) of cosmic-ray particles to be identified. It consists of 768 silicon micro-strip sensors assembled in 6 double layers with a total active area of \totalarea~m$^2$. Silicon planes are interleaved with three layers of tungsten plates, resulting in about one radiation length of material in the tracker. Internal alignment parameters of the tracker have been determined on orbit, with non-showering protons and helium nuclei. We describe the alignment procedure and present 
 the position resolution and alignment stability measurements. 
\end{abstract}

\begin{keyword}
Cosmic-Ray Detectors, Gamma-Ray Telescopes, Alignment, Silicon-Strip Detectors.
%\texttt{elsarticle.cls}\sep \LaTeX\sep Elsevier \sep template
%\MSC[2010] 00-01\sep  99-00
\end{keyword}

\end{frontmatter}

%\linenumbers

\section{Introduction}

\begin{figure}[]
\includegraphics[width=0.55\textwidth]{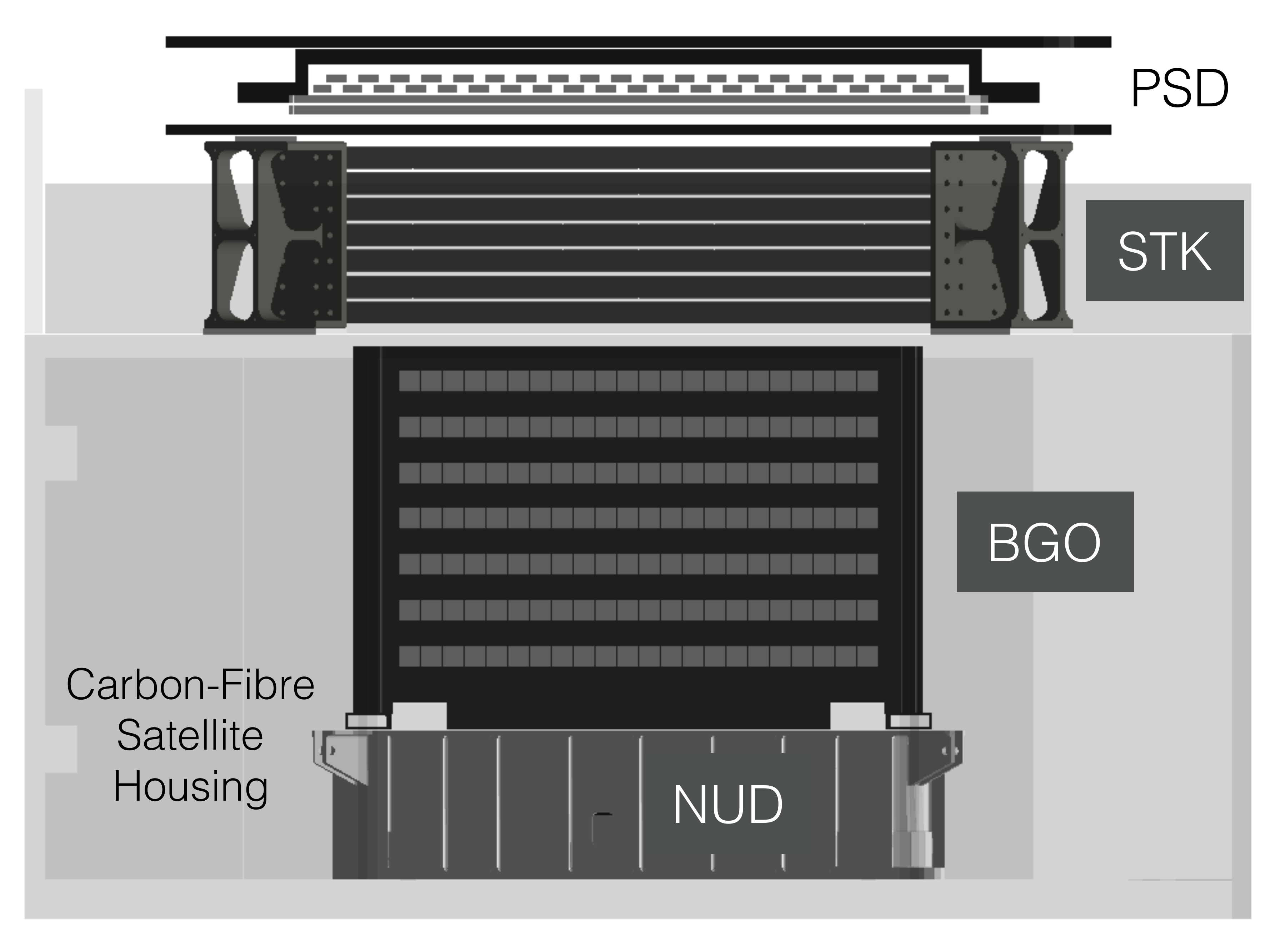}
\includegraphics[width=0.50\textwidth]{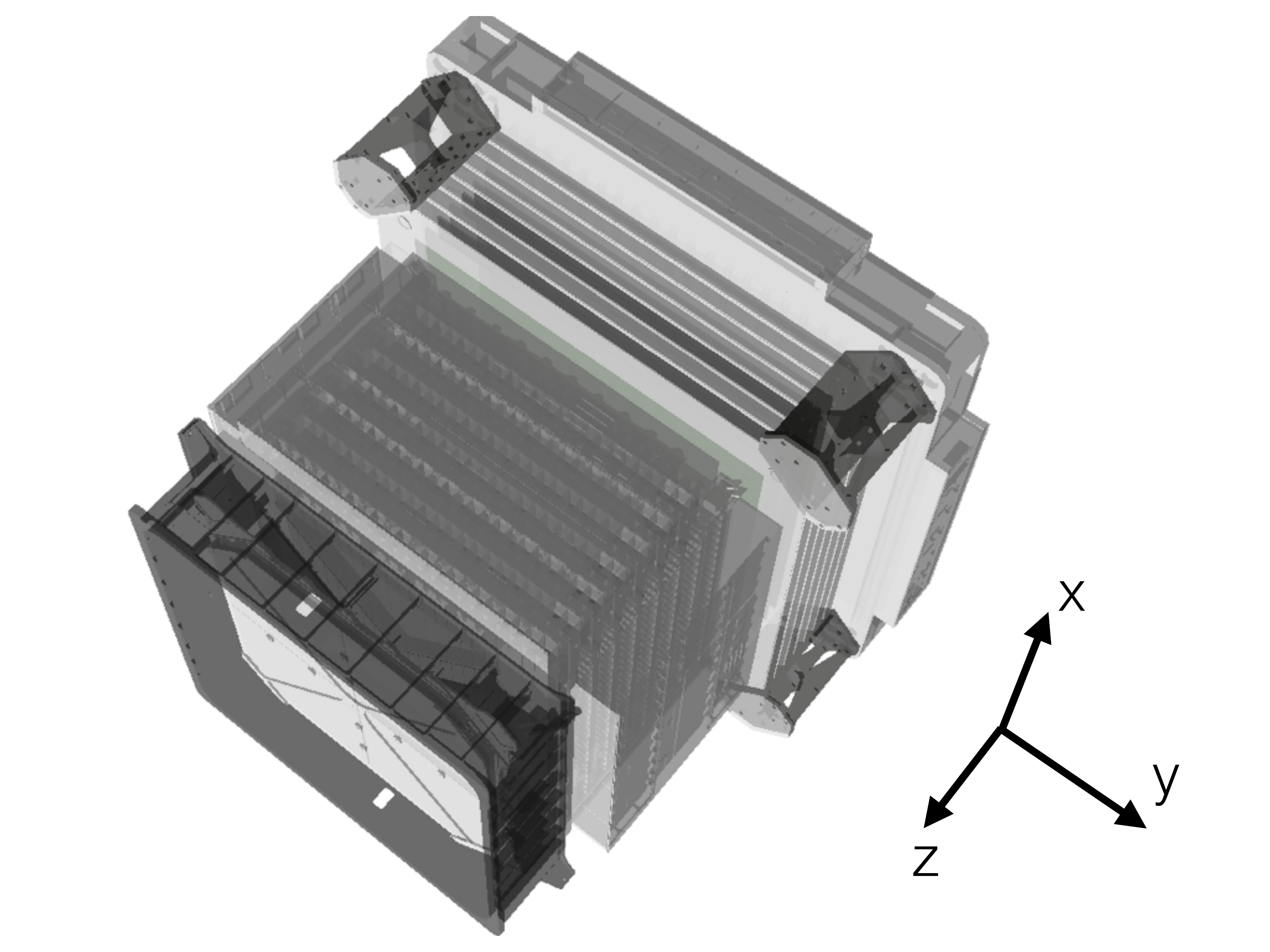}
\caption{Schematic view of the DAMPE detector. Sensitive detectors and support structures are shown. The $z$-axis of the DAMPE coordinate system is oriented to the zenith, orthogonal to the STK planes and $y$ points to the Sun.
}
\label{fig:dampe}
\end{figure}

The DArk Matter Particle Explorer (DAMPE) is a satellite mission that was launched on December 17, 2015 from the Jiuquan Satellite Launch Center in the Gobi Desert, China. It is designed to detect electrons and photons in the few GeV to \mbox{10 TeV} energy range, as well as  protons and cosmic-ray ions from \mbox{10 GeV} to \mbox{100 TeV}, with excellent energy resolution and direction precision~\cite{Chang:dampe,TheDAMPE:2017dtc}. The main objectives of DAMPE are the identification of possible indirect signatures of Dark Matter annihilation or decay, improving the understanding of the origin and propagation mechanisms of high energy cosmic rays and gamma-ray astronomy. It consists of four sub-detectors (Figure~\ref{fig:dampe}) stacked together as follows, moving from top to bottom. First is a Plastic Scintillator-strip Detector (PSD), which measures the cosmic ray charge (Z) and provides the veto signal for charged particles in gamma-ray detection. It is followed by a Silicon-Tungsten tracKer-converter (STK), that is described in detail in the next section. Next, there is an imaging calorimeter made of 14 layers of Bismuth Germanium Oxide (BGO) bars in a hodoscopic arrangement with a total thickness of about 32 radiation lengths, which provides a precise energy measurement and particle identification for electron/hadron separation. The BGO is aided by the NeUtron Detector (NUD), a boron-doped plastic scintillator detecting delayed neutrons coming from hadronic interactions at high energies, which serves to improve the electron/hadron separation power. %~\cite{Chang:dampe,TheDAMPE:2017dtc}. 

  The STK is a key component of DAMPE, allowing the trajectory and absolute ion charge (Z) of incoming particles to be reconstructed and measured respectively. Moreover, thanks to its high position resolution, the direction of incoming photons converting into electron-positron pairs in the STK's tungsten plates can be precisely reconstructed.  In order to fully exploit the trajectory reconstruction capabilities of the STK, a precise alignment of the instrument is needed, as explained in this paper. 

The paper is organized as follows. In Section~\ref{sec:STK} the STK is briefly described. Section~\ref{sec:data} provides the details of the on-orbit data and simulation used in the alignment analysis. Section~\ref{sec:data_processing} gives an overview of the data reconstruction procedure. In Section~\ref{sec:alignment_analysis} the alignment procedure is described in detail. In Section~\ref{sec:resolution} the results on the STK position resolution are reported. In Section~\ref{sec:stability} the alignment stability and its on-orbit variations are discussed. Conclusions are  given in Section~\ref{sec:conclusion}.

\section{The STK}
\label{sec:STK}

\begin{figure}[]
\includegraphics[width=0.5\textwidth]{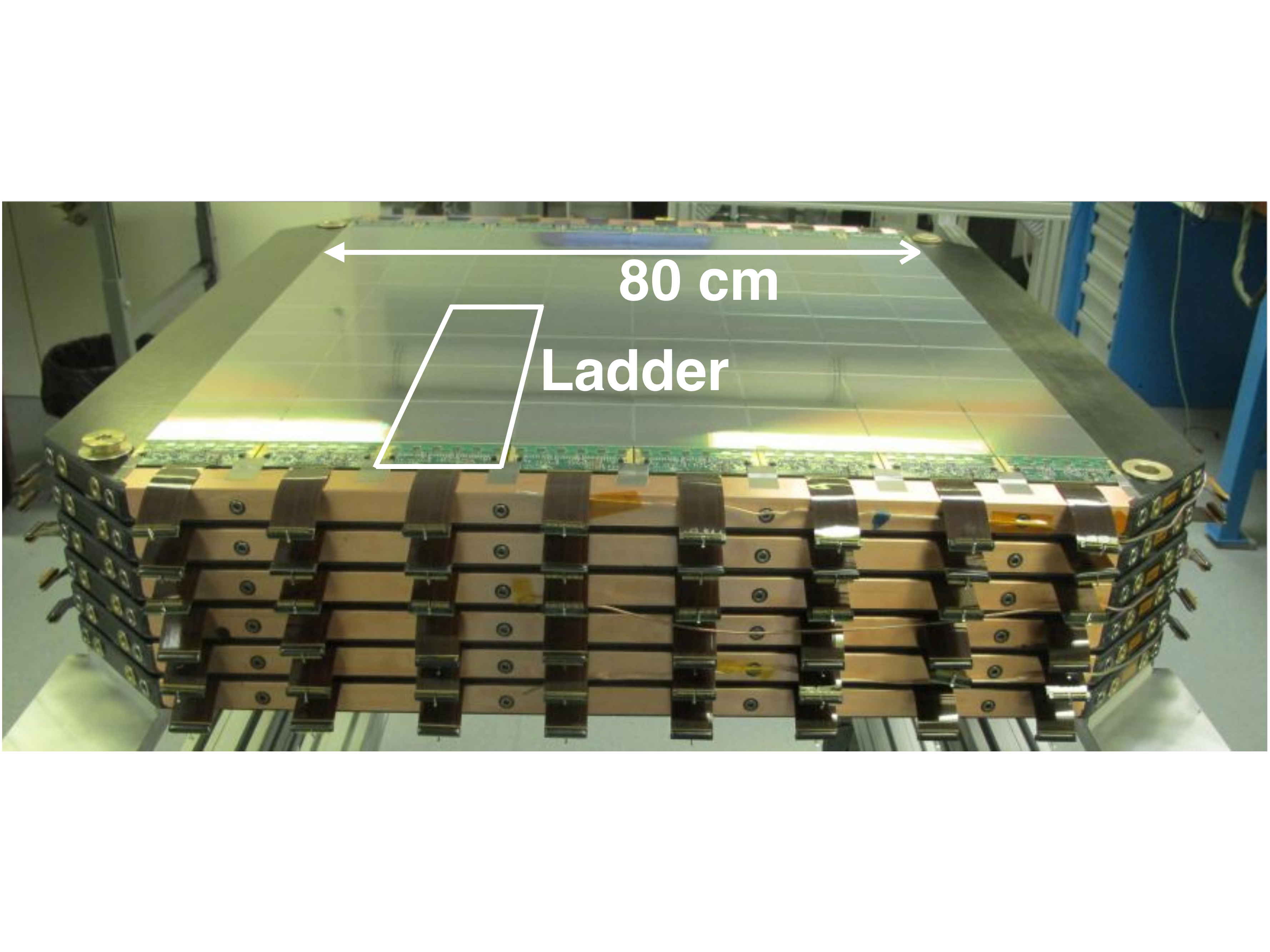}
\includegraphics[width=0.5\textwidth]{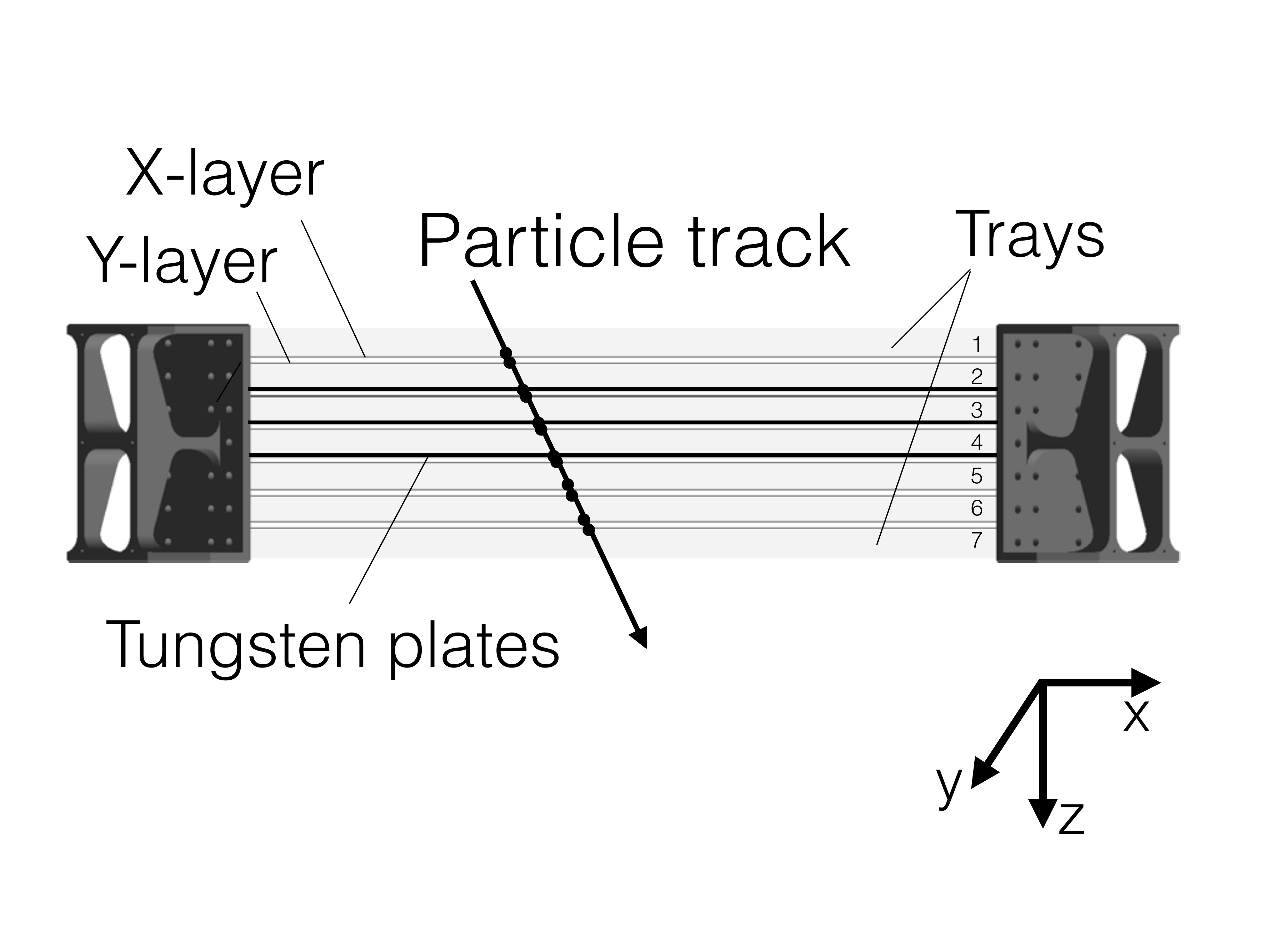}  % Matteo
\caption{(Left) A picture of the STK before the assembly of  the last tray. (Right) Schematic view of the STK with the following parts highlighted: the silicon layers (gray), the tungsten converters (black), the aluminum corner feet (black) and the supporting trays of carbon fiber and aluminum-honeycomb (pale gray).}
\label{fig:stk}
\end{figure}

The STK~\cite{Azzarello:2016trx} is designed to reconstruct the charged particle trajectories, to identify the direction of incoming gamma-rays converting into electron-positron pairs and to measure the charge~$\mathrm{Z}$~of cosmic rays.  It consists of 6 tracking double-layers, providing 6 independent measurements of the $x$ and $y$ coordinates of the incoming particle. The tracking layers are mounted on 7 supporting trays, as shown in Figure~\ref{fig:stk}. To favor the pair conversion of incoming photons into electron-positron pairs, three tungsten layers are placed after the first, second and third tracking layer. Each tungsten layer is 1 mm thick, for a total of about one radiation length.

Each silicon layer consists of 16 modules, hereafter referred to as  ladders (Figure~\ref{fig:stk} left). Each ladder is formed by 4 single-sided AC-coupled Silicon micro-Strip Detectors (SSD), daisy-chained via  micro-wire bonds. All the 192 ladders are readout in groups of 24 by 8 data acquisition boards~\cite{Zhang:2016hth}. The sensors are produced by Hamamatsu Photonics~\cite{hamamatsu} and are 320~\micron~thick, with dimensions of $\mathrm{9.5}$ $\times$ $\mathrm{9.5}$ $\mathrm{cm}^2$. They are segmented into 768 strips with a 121~\micron~pitch, with a sensitive SSD area of $\mathrm{9.29}$ $\times$ $\mathrm{9.29}$ $\mathrm{cm}^2$. 
 The readout is done for every other strip, in order to limit the number of readout channels whilst maintaining sufficient spatial resolution~\cite{gallo}. 
The resulting 384 channels per ladder are read out by 6 VA140 ASIC chips made by IDEAS~\citep{ideas}. With analog readout and charge sharing on the non-readout strips, the expected position resolution is better than 70~\micron~for most incident angles.

\begin{figure}[]
\begin{centering}
\includegraphics[width=1.0\textwidth]{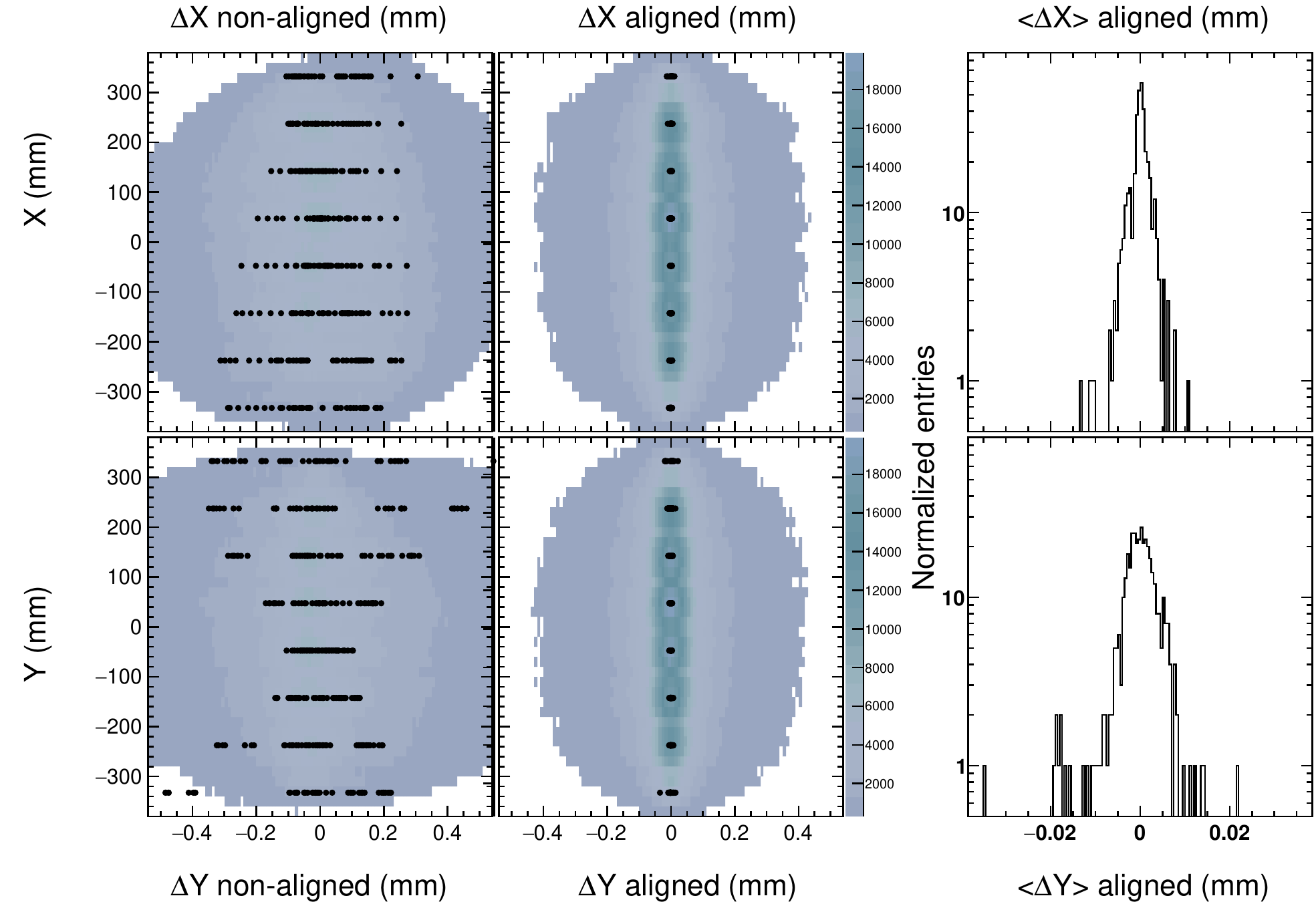}
\caption{Track $x$- and $y$-residuals before (left) and after (middle) the alignment procedure, shown with respect to the particle coordinates. 
 Black circles represent the average residuals for each of the 768 sensors with respect to the nominal  sensor position, defined as the center of the SSD: 
384 for $x$ (top) and 384 for $y$ (bottom). The distributions of these values are reported in the plots on the right side, where each histogram entry corresponds to one SSD.}
\label{fig:align_example}
\end{centering}
\end{figure}

The mechanical assembly of the STK has a construction precision of about 100~\micron, coarser than the expected position resolution of the silicon sensors. 
Hence, alignment parameters are introduced to correct for the displacement and rotation of each sensor with respect to its nominal position, allowing the full tracker potential to be exploited. 
 The alignment parameters are derived from the data, and they allow the correction of the particle coordinate in the STK, as described in Section~\ref{sec:alignment_analysis}. The effect of the alignment procedure is illustrated in Figure~\ref{fig:align_example}. The left and middle plots show the  residual values with respect to the particle coordinate, before and after the alignment respectively. The residual is defined as the difference between the measured coordinate and the projection of the track in the \emph{i}-th plane where the track is reconstructed without considering the \emph{i}-th plane. Plots on the right show the distributions of the average residuals of SSDs. For these plots \mbox{1.3~$\mathrm{M}$} events passing the event selection described in  Section~\ref{sec:event_selection} are used.

\section{Data and simulation}
\label{sec:data}
\subsection{Data}

The DAMPE satellite  is operating at an altitude of about 500 km in a sun-synchronous orbit with an inclination of about 97$^{\circ}$, in a sky-survey mode permanently oriented to the zenith. Each orbit lasts about 95 minutes~\cite{TheDAMPE:2017dtc}.  DAMPE has operated smoothly since its launch, recording about five million events per day. The data used in the  analysis presented here was collected between January 2016 and May 2017.

\subsection{Simulation}

Monte-Carlo simulations of the full detector are used to validate the alignment procedure. Proton and helium events are generated based on the GEANT4~\cite{Agostinelli:2002hh} toolkit, following a power-law energy spectrum with an index of -1. The energy range and statistics of the Monte-Carlo samples are reported in Table~\ref{tab:mc_samples}.

 \begin{table}
 \begin{center}
 \begin{tabular}{c r@{--}l c}
\hline
Generated Particle Type & \multicolumn{2}{c}{Energy Range} & Statistics (events) $\times10^6$ \\
\hline
$p$ & 1 GeV~~&~100 GeV & 500 \\
$p$ & 100 GeV~~&~10 TeV & 100 \\
$\mathrm{He}$ & 10 GeV~~&~100 GeV & ~20 \\
%$\mathrm{He}$ & 50 GeV~~&~100 GeV & ~20 \\
$\mathrm{He}$ & 100 GeV~~&~10 TeV & 100 \\
\hline
\end{tabular}
\caption{Monte-Carlo samples used for the alignment analysis.}
\label{tab:mc_samples}
\end{center}
\end{table}

 The detector geometry, including all sensitive volumes and the supporting structures of the payload, is implemented in the simulation from Computer-Aided-Design (CAD) drawings, using a software conversion toolkit~\cite{cad-to-gdml,tykhonov_icrc,Wang:2016yov}.  To account for the effect of capacitive charge sharing between individual silicon strips,  the sharing coefficients were obtained from a detailed SPICE~\cite{spice} simulation of the SSD-equivalent electronic circuit. For a fair comparison with the events collected during on-orbit operation, the simulated data is processed through the same reconstruction and analysis chain as real data.

\section{The STK event reconstruction}
\label{sec:data_processing}

The reconstruction algorithms of the STK signal can be  grouped into  two main procedures: hit reconstruction and track finding.  

\subsection{Hit reconstruction}

The raw data of the STK are represented as 12-bit ADC values, for a total of 384 values (channels) per ladder. %The ADC values are processed in an FPGA
 The ADC values are processed on-board by FPGAs, to reduce the data size~\cite{Dong:2015qma}. The data compression algorithm  performs a simple clustering of the signals, where clusters are formed starting from channels (seeds) exceeding the signal-to-noise ratio $S/N >3.5$, and including all neighboring channels  with $ADC > 5$.

A more sophisticated clustering is then performed offline. The ADC values are grouped into arrays of 384 channels for each ladder.   The clustering algorithm scans the arrays to find seeds, which are defined as local signal maxima with $S/N > 4$,  then clusters are formed by collecting all neighboring  channels with $S/N >1.5$. If the signal in two neighboring channels of the cluster, while moving away from the seed channel,  increases more than five times the corresponding average noise for these channels, the cluster is considered to have multiple signal peaks and is split into two clusters. High-noise or broken channels (which are less than 1\% of the total) are masked during the reconstruction. The cluster coordinate is measured in units of readout channels, and calculated as the Center-of-Gravity (CoG) of the cluster strips weighted by the ADC signals, defined as 
\begin{equation*}
\mathrm{CoG} =\frac{  \sum_{i}{i\cdot \frac{S_{i}}{N_{i}}}  }      {    \sum_{i}{\frac{S_{i}}{N_{i}}} }
\end{equation*}
where the sum is taken over all channels of the cluster. $i$ is the index number of a channel and  $S_{i}$  ($N_{i}$) is the signal (noise) value of the $i$-th channel. The average cluster size for protons is found to range from 2.3 to 2.9 channels, depending on the arrival direction. For helium nuclei the average cluster size varies in the range of 3.2 to 4.0 channels. 

The $x$ and $y$ clusters in the same tracking plane are then combined in all possible permutations to form 3D hits. Since each $xy$ quarter plane is read out by different  electronic boards, only combinations coming from the same quarter plane are allowed, which significantly reduces the number of false hits due to combinatorics. % possible hits.

The hit coordinate in the global reference frame of the detector is obtained from the  cluster CoG, the ladder position, $x_{L}$, and the ladder orientation  in the nominal design, $R_{L}$, as follows:
\begin{equation*}
x=\mathrm{CoG}\cdot d \cdot R_{L}+x_{L}.
\end{equation*}
where $d=242$~\micron~ is the strip readout pitch; $R_{L}$ is equal to $\pm1$, depending on the channel numbering orientation in the ladder readout.   % (distance between two read-out strips of a silicon sensor). 
The same holds for $y$.

\subsection{Track finding}

To reconstruct a track, we start from the shower direction of the BGO obtained with a centroid approach. This direction is projected to the first STK layer (the closest to  the PSD) to select the closest hit. The track seed is then formed from this hit and the shower direction and the track is reconstructed using a custom implementation of the Kalman filter algorithm~\cite{Fruhwirth:1987fm}. The quality of the track is evaluated from the number of hits and the $\chi^2/\mathrm{ndof}$ of the Kalman fit. Track candidates are rejected if the $\chi^2/\mathrm{ndof}$ is above a threshold that depends on the number of hits. The initial point of the inspected track is removed from the list of the seed points, and the seeding-filtering is repeated with the next-to-closest hits in the first layer. 
 
 Once all the seed points in the first layer have been used, the procedure is repeated with the layers 2 and 3. In this case, after the track is reconstructed, an additional step takes place as follows. The hits are searched in the layers before the seeding-hit plane using the track direction, and, if matching hits are found, they are included into the track, which is refitted with the appended hits. The seed-filtering is repeated until all seed points have been exhausted.  
 
Finally, the above procedure is repeated with the three layers closest to the BGO (layers 6, 5 and 4) with the inverse direction of track filtering (from the BGO to the STK).  On average, 41\% of reconstructed tracks are recovered with the inverse track finding.
%The average fraction of tracks  in the data reconstructed using the inverse track-finding direction is 40\%. 

\section{The STK alignment}
\setcounter{footnote}{0}
\label{sec:alignment_analysis}

\subsection{Event selection}
\label{sec:event_selection}

\begin{figure}[]
\begin{centering}
\includegraphics[width=1.0\textwidth]{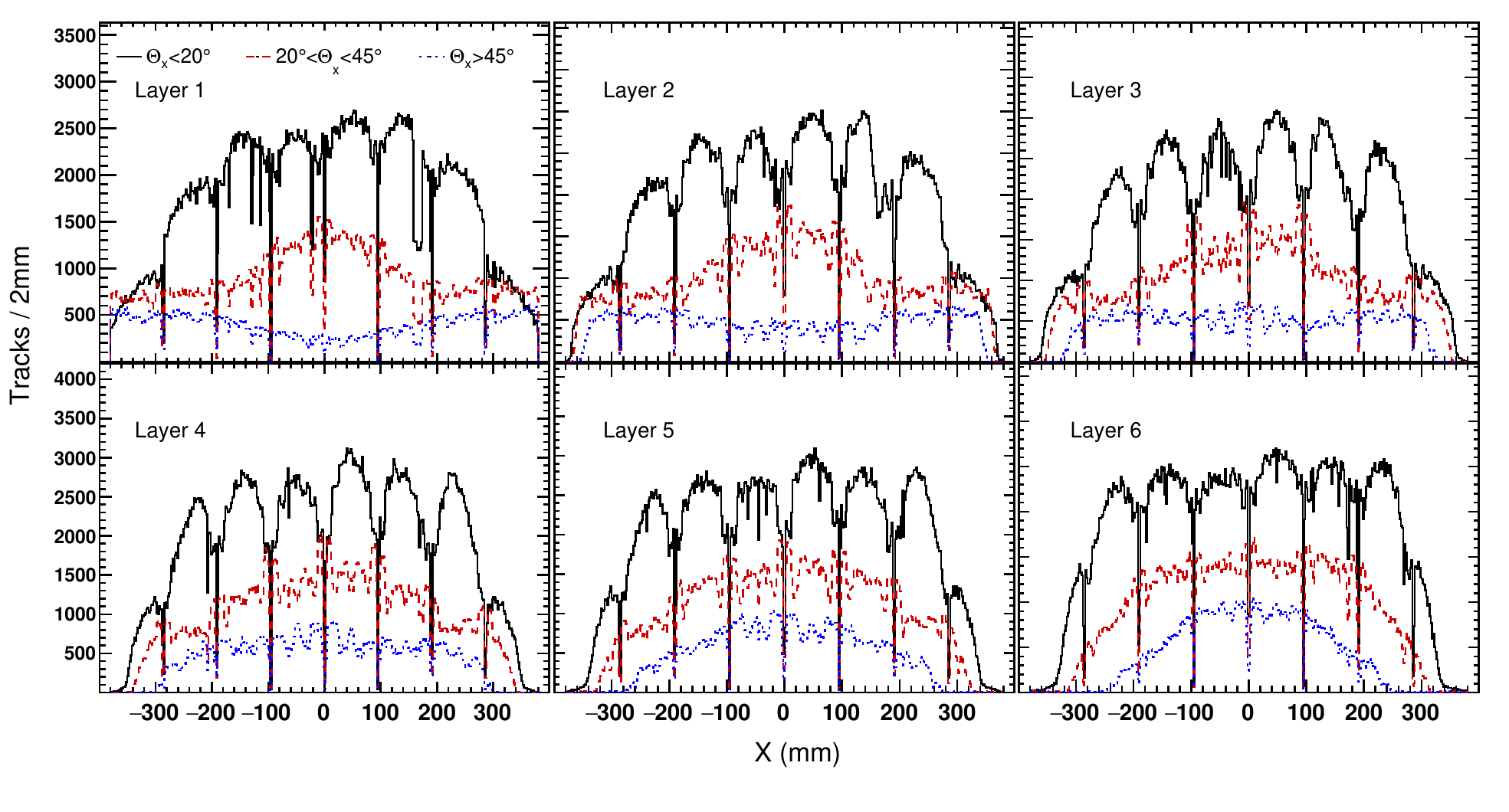}
\caption{Distributions of tracks in the alignment data sample with respect to the $x$ position of track in different layers of the STK, shown for different track inclinations: $\mathrm{\theta_x < 20^\circ}$ (solid line), $\mathrm{20^\circ<\theta_x < 45^\circ}$ (dashed line) and $\mathrm{\theta_x > 45^\circ}$ (dotted line). Statistics in the plot corresponds to about 1.5 days of data after the alignment event selection.  Layer 1 corresponds to the STK layer closest to the PSD. The structure of the equidistant drops   is due to particles passing in the dead regions of the silicon sensors.}
\label{fig:luminocity}
\end{centering}
\end{figure}

Events used in the alignment analysis are required to contain exactly one track, where the track is required to have $x$ and $y$ hits in all six planes.  Not only does this simplify the involved derivatives calculation associated with the gradient-descent alignment method, but it also helps to minimize the track projection error in the estimate of the position resolution. 
 About \aligneventss\% of  events in the full data set pass this selection criterion, which corresponds to about \mbox{\eventsforalignmentdaily~events} per day.  About 80\% of events do not have tracks passing this selection, and 5\% of events have more than one track. These events are not used in the analysis.  For the alignment procedure, reconstructed tracks are refitted with a straight line, using a simple least squares method, both in $x$--$z$ and $y$--$z$ projections independently. Tracks with $\chi^2$-values higher than a certain pre-defined threshold are removed from the sample\footnote{The $\chi^2$ is evaluated for a linear fit of non-aligned tracks, as follows $\chi^2=\sum_{i}(x_i^{fit} -x_i^{hit})^2+\sum_{i}(y_i^{fit} -y_i^{hit})^2$, where the sum is taken over all points comprising the track. Afterwards, tracks which have $\chi^2$ higher than the mean value plus 5 times the RMS of the $\chi^2$ distribution are removed from the sample.}. About 98\% of the events pass this selection.  Figure~\ref{fig:luminocity} shows the distributions of tracks with respect to the $x$ coordinate in different STK layers after the alignment event selection.

\subsection{Alignment algorithm}

The following alignment correction parameters are introduced for each silicon sensor of the STK:
\begin{equation}
\Delta_X / \Delta_Y, \quad \Delta_Z, \quad \theta_{X}, \quad \theta_{Y}, \quad \theta_{Z}.
\end{equation} 
They correspond to the two offsets and three rotations of the silicon sensor respectively, for a total of 3840 alignment parameters. Rotations are considered with respect to the origin of the $xy$ reference frame in the plane where the sensor resides in the nominal design (CAD detector geometry).  Bending of SSDs may contribute up to 1~\micron~to the position resolution at high particle incidence angles, according to  the STK  ladder metrology data. However, while we originally considered including the bending parameters of the SSDs as additional free parameters, we found it to have no significant impact on the resulting position resolution of the STK, and thus we treat each silicon sensor as a rigid body. The corrected (aligned) position of  a hit can then be written as set of coupled equations:
\begin{align}
x_{a} &= x + \Delta_X - y \cdot \theta_{Z} \\
y_{a} &= y + \Delta_Y + x \cdot \theta_{Z} \\
z_{a} &= z + \Delta_Z - x \cdot \theta_{Y} + y \cdot \theta_{X}
\end{align}
The method used to determine the alignment parameters is based on the optimization (minimization) of the total $\chi^2$-value of tracks in the alignment data sample, where the total $\chi^2$  is defined as follows: 

\begin{equation}
\chi^2 = \sum_{t\in\{\mathrm{tracks}\}}{\left(
\sum_{p\in\{\mathrm{points}\}}{     \frac{ \left(x_{t,p}^{fit}-x_{t,p}^{hit} \right)^2 }  {N_{\mathrm{track},s_x}   } } +
\sum_{p\in\{\mathrm{points}\}}{    \frac{ \left(y_{t,p}^{fit}-y_{t,p}^{hit} \right)^2 }   {N_{\mathrm{track},s_y}  } } 
\right)},\\
\label{eq:chisq_track_2}
\end{equation}
\begin{align}
s_x = \mathrm{s_x(t,p)},~s_y = \mathrm{s_y(t,p)}.
\end{align}
where $t$ is the index of track in the sample, $p$ is the index of the point within the track, $s_x$  and $s_y$  are the SSD identifiers, $x^{hit}$ and $x^{fit}$ are the measured and fitted (from a linear fit) coordinates of the track in the $i$-\emph{th} point and $N_{\mathrm{track,}s_x/s_y}$  is the number of tracks crossing the corresponding SSD. The $N_{\mathrm{track,}s_x/s_y}$  factor allows the non-uniformity of track statistics for different SSDs to be taken into account.  Note that alignment can not be done in $x$ and $y$ independently, since both coordinates are needed when identifying which SSD a track hit is associated with.

The minimization is done through an iterative gradient descent procedure using the gradient of $\chi^2$  with respect to the alignment parameters. In this way, all alignment parameters are optimized simultaneously. Specifically, the movement in the phase space of the alignment parameters is performed in the opposite direction to the gradient.  After each iteration, the gradient is recalculated and another step is performed. The step size of the gradient descent varies dynamically. 

Two custom implementations of the gradient descent method were tested in the analysis, as described below.

\subsubsection{The~\fixedchisqare~method}

In this method, the $\chi^2$ value for each iteration of the algorithm is compared with the one from a previous iteration, and if the improvement falls below a pre-determined threshold, the step size is halved. The condition at which the step size is decreased is:
\begin{equation}  
  \chi^2_{next} \geq \chi^2_{this}-\epsilon \cdot  \left\|  \vec{\bigtriangledown}\chi^2 \right\|  \cdot  \left\|  \vec{d} \right\| 
\end{equation}
where $\vec{\bigtriangledown}\chi^2$ and $\vec{d}$ are the gradient and the step vectors respectively and $\epsilon \in [0;1]$ is the threshold parameter. We tested three different possible values for this threshold: $0$, $0.5$ and $1$. In the case of $\epsilon=0$, the algorithm oscillates around the minimum of the objective function (total $\chi^2$) and converges slower than in the case of~$\epsilon=0.5$. However, in the case $\epsilon=1$ the step size of the algorithm drops after only a few iterations and therefore the algorithm does not converge. As a result, we chose $\epsilon=0.5$  as a reference threshold value. We refer to this as the~\fixedchisqare~method, since the set of tracks for the $\chi^2$ evaluation remains fixed at each iteration of the alignment algorithm.

\subsubsection{The~\variablechisqare~method}

The alignment precision is limited by the internal resolution of the silicon sensors, and more significantly by the presence of multiple scattering (MS). To reduce the MS contribution, we apply here a further selection on the track candidates, rejecting tracks with track-hit residual values above a defined threshold.  Specifically, we reject tracks if at least one track-hit residual is higher than~\mbox{300~\micron}.
 This selection is applied on top of the track selection described in Section~\ref{sec:event_selection}. % used for the~\fixedchisqare~algorithm.  
 The~\mbox{100~\micron}~threshold was also tested and no significant difference was found with respect to the ~\mbox{300~\micron} one. 
 Furthermore,  we allow the track sample to change from one iteration to another, so that the residual-based track quality selection is applied at every iteration of the alignment, introducing some new tracks in the sample and (or) removing some old ones. Contrary to the~\fixedchisqare~minimization, in this approach $\chi^2$ values cannot be used to control the step size of the algorithm. Hence we set a requirement on the angle between gradient vectors for two subsequent iterations, ~$\alpha$, as follows: %\emph{similarity} of the gradient vectors for two subsequent iterations to be larger than $\epsilon=0.5$, as follows:
 \iffalse
\begin{equation}
\mathrm{cos}\alpha\left( \vec{\bigtriangledown}\chi^2_{this}, \vec{\bigtriangledown}\chi^2_{next}\right) \equiv
\frac{\vec{\bigtriangledown}\chi^2_{this}\cdot \vec{\bigtriangledown}\chi^2_{next}}{        \left\|  \vec{\bigtriangledown}\chi^2_{this}  \right\|     \left\|  \vec{\bigtriangledown}\chi^2_{next}  \right\|        } > \epsilon
\end{equation}
\fi
\begin{equation}
\mathrm{cos}(\alpha) \equiv
\frac{\vec{\bigtriangledown}\chi^2_{this}\cdot \vec{\bigtriangledown}\chi^2_{next}}{        \left\|  \vec{\bigtriangledown}\chi^2_{this}  \right\|     \left\|  \vec{\bigtriangledown}\chi^2_{next}  \right\|        } > \epsilon
\end{equation}
where $\epsilon=0.5$. When the cosine is less than $\epsilon$ the step size is halved. We call this approach the~\variablechisqare~ method.

\begin{figure}[]
\begin{centering}
\includegraphics[width=0.8\textwidth]{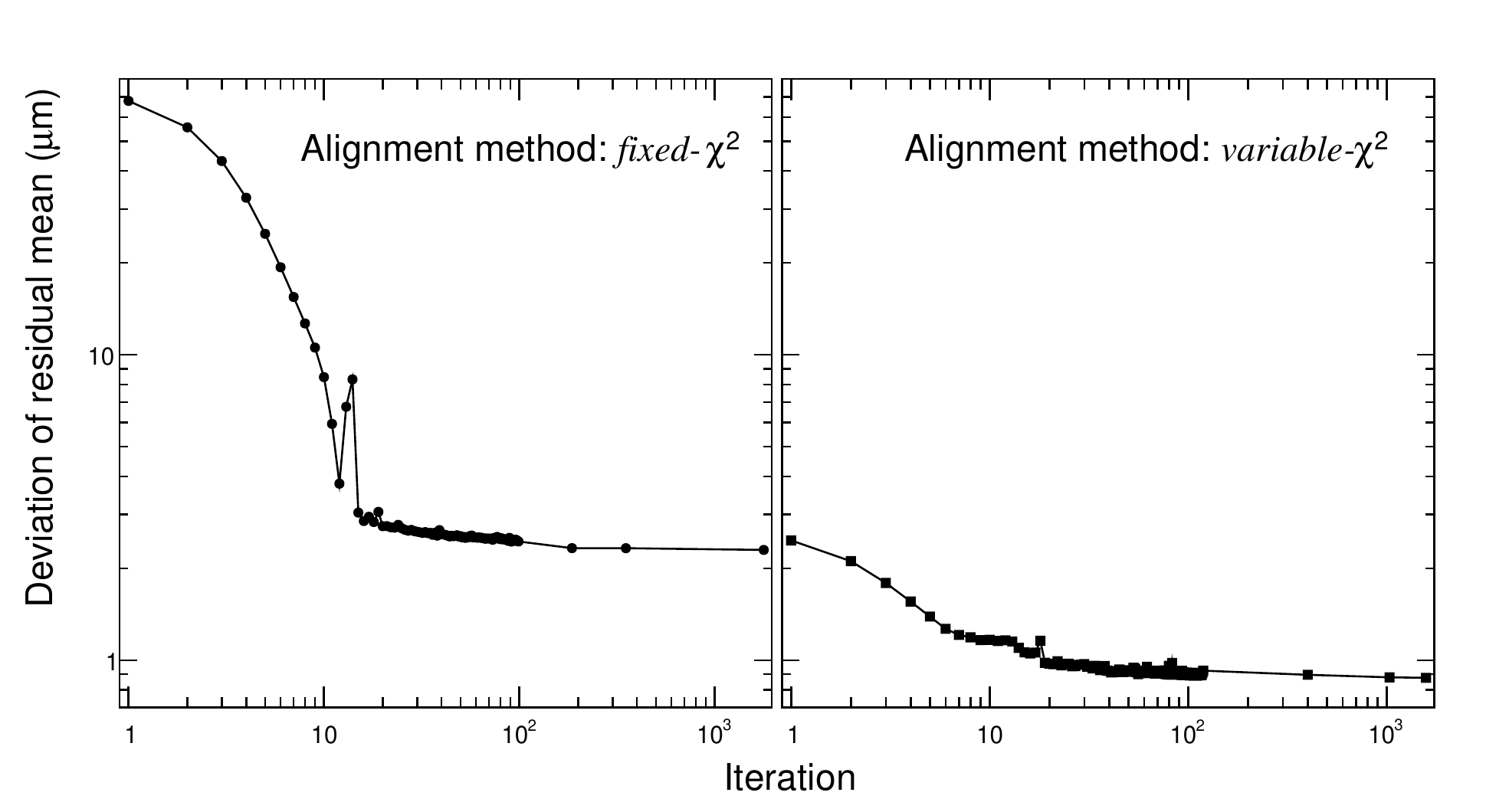}
\caption{The \emph{deviation of residual mean} for track-hit residuals after the alignment as a function of the number of iterations of the alignment algorithm for~\fixedchisqare~(left) and~\variablechisqare~(right) methods.  The peak around 10 iterations in the left plot is due to additional iterations that are needed because of the gradient-descent step becoming too big and missing the minimum of the objective function.}
\label{fig:alignment_iterations}
\end{centering}
\end{figure}

Figure~\ref{fig:alignment_iterations}~shows the comparison between the~\variablechisqare~and the~\fixedchisqare~methods. The \emph{deviation of residual mean} is defined as the RMS of the distribution  of average residuals for all 768 silicon sensors.  As shown in the figure the~\variablechisqare~method performs better  than the~\fixedchisqare~one, yielding a lower deviation. Therefore the~\variablechisqare~is used as the baseline alignment method  for the remainder of the paper.

\begin{figure}[]
\begin{centering}
\includegraphics[width=0.86\textwidth]{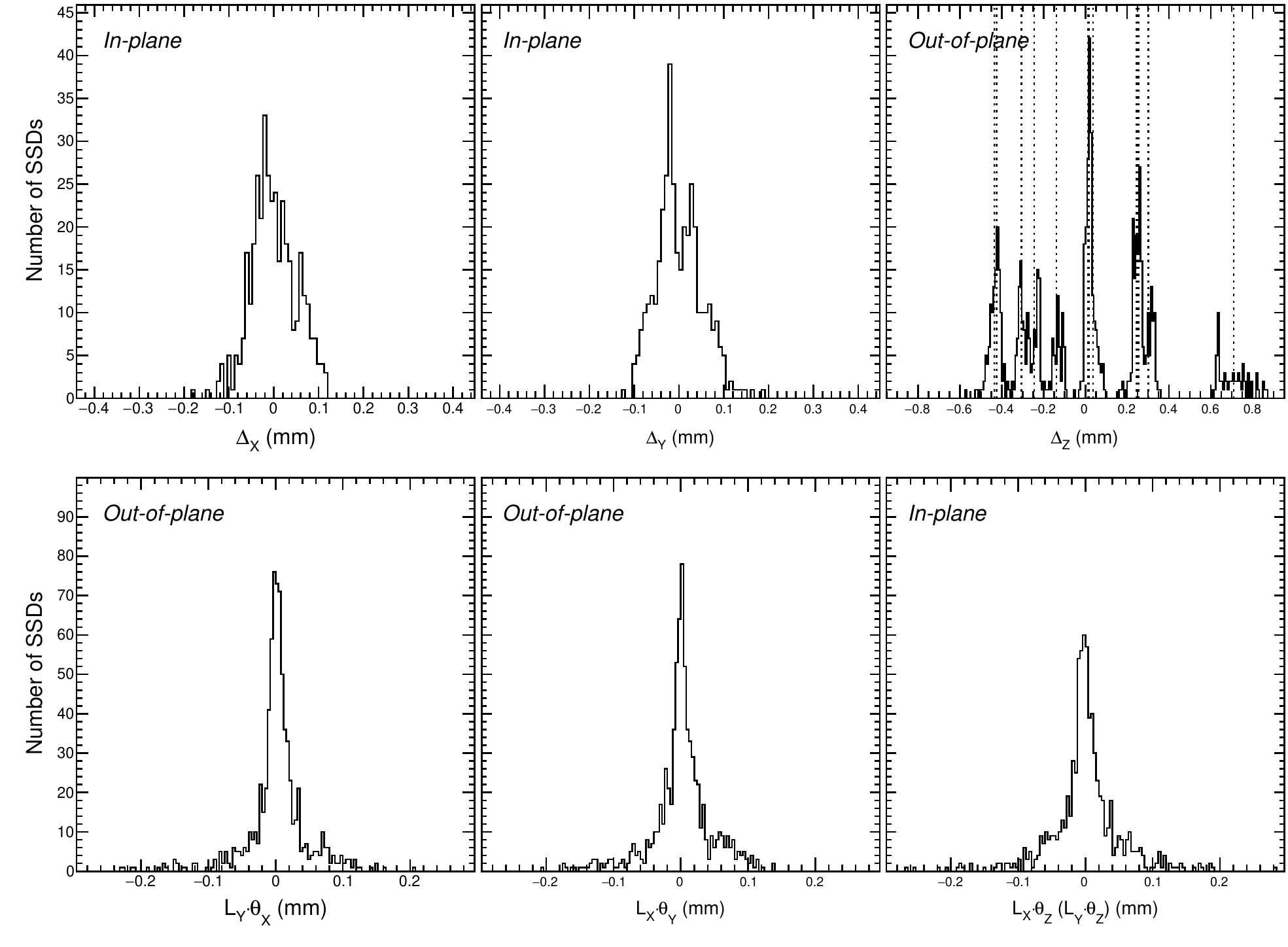} 
\caption{Distributions of the alignment parameters for all the silicon sensors. For compatibility with the offsets, rotation parameters are multiplied by the corresponding positions of silicon sensors (lever arms of rotation). The $z$ shifts and rotations around $x$ and $y$ axes are indicated as \emph{out-of-plane}. For the top right plot, the peaks structure is due to the different STK planes. Dotted lines indicate the average z shifts for each  of 6 $x$ and of 6 $y$  layers of the STK.}
\label{fig:all_alignments}
\end{centering}
\end{figure}

The first on-orbit alignment of the STK was performed with the data collected during two consecutive days in January 2016. Distributions of the obtained alignment parameters for all the silicon sensors are shown in Figure~\ref{fig:all_alignments}. The distributions for $\Delta_X$ and $\Delta_Y$ parameters are centered around zero with a width of about \mbox{100~\micron}, as expected from the mechanical precision. For the $\Delta_Z$ parameters, the structure of the $z$ shifts corresponding to 6 different tracking planes can be observed.

\section{Position resolution}
\label{sec:resolution}

The position resolution of the STK is estimated from the residual distributions of proton and helium candidate tracks after applying the alignment. The residuals are defined as the difference between projected and measured hit position, where the projected position is obtained from the linear fit of the remaining points of the track, without the point being tested.

\subsection{Position resolution with protons}
\label{sec:resolution_protons}

We apply a further event filtering to select high-energy tracks and reduce the contribution of multiple scattering to the residuals. Events are required to fulfill the following criteria:
\begin{itemize}
\item the total reconstructed energy in the calorimeter~$E>50$~GeV;%is above 50 GeV;  
\item the angular distance between the calorimeter shower direction and the track direction is below 10 degrees; %0.2 radians; %\emph{rad};
\item the distance between the CoG of the calorimeter shower and the track projection in the BGO  is below 50 mm, where the distance is measured in the $x$-$y$ plane passing through the CoG of the shower.
\end{itemize}

\begin{figure}[]
\begin{center}
\includegraphics[width=0.85\textwidth]{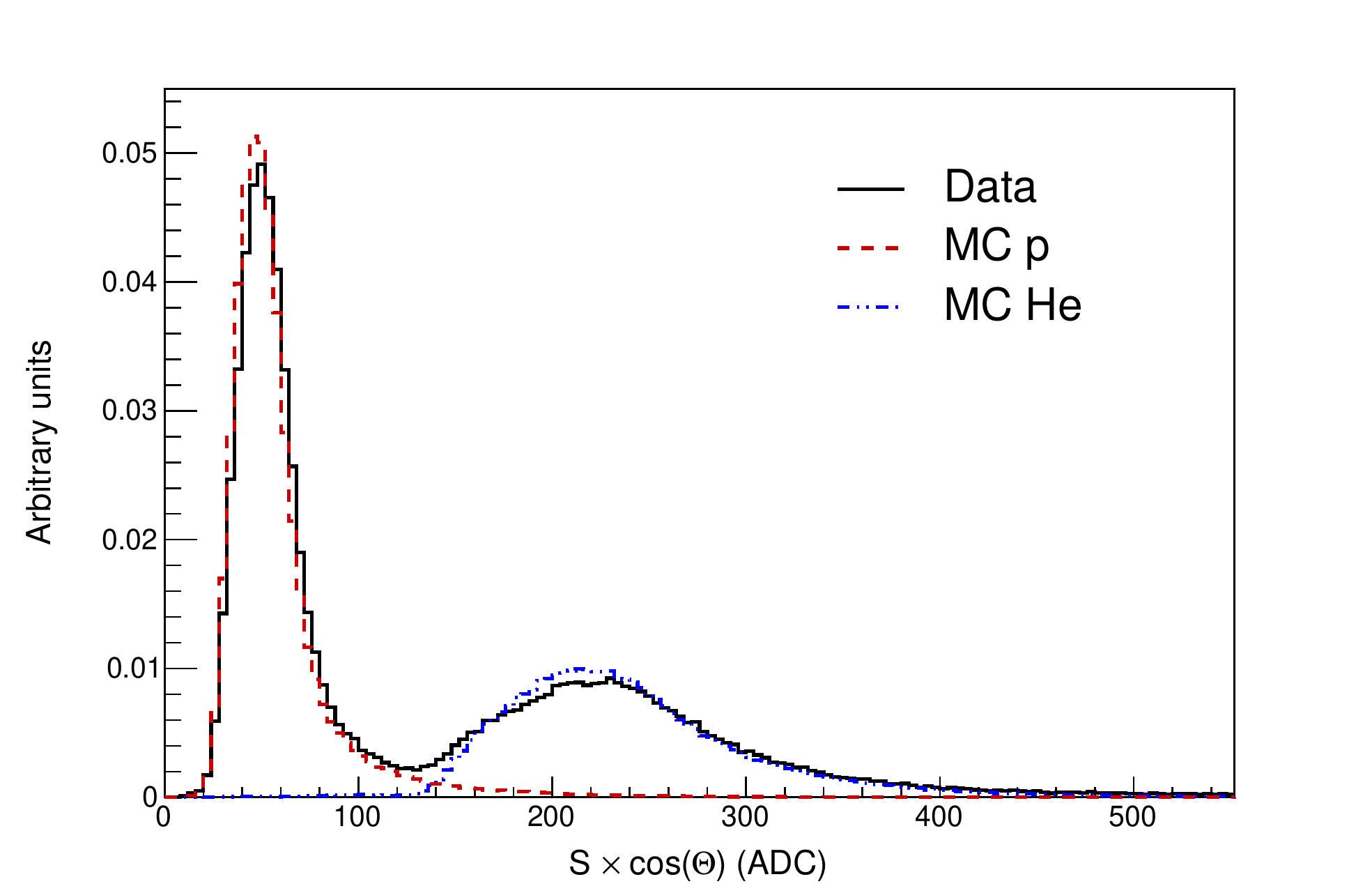}
\end{center}
\caption{The normalized distribution of all hits in the tracks with respect to the hit cluster charge multiplied by the cosine of the track incidence angle. The tracks are required to pass the selection described in Section~\ref{sec:resolution_protons}.}  
\label{fig:p_he_selection}
\end{figure}

Moreover, in order to select protons, a cut on the cluster charge of each hit associated with the track is applied.  The cluster charge, corrected for the particle path in the detector,~\hitsignal, is required to be lower than 160 ADC counts. If at least one hit in the track does not pass this selection criterion, the track is discarded. As an example, Figure~\ref{fig:p_he_selection} shows the cluster charge distribution for all the track hits before the cluster ADC selection is applied. The cluster ADC cut is introduced to eliminate the contribution of helium and heavier ions in the sample, and has a 78\% efficiency for protons,  with a residual helium contamination of about 0.1\%, estimated from the Monte-Carlo simulation.  About~\mbox{0.15~$\mathrm{M}$}~events per day in the data pass the aforementioned cosmic-ray proton selection criteria.

\begin{figure}[]
\includegraphics[width=1.0\textwidth]{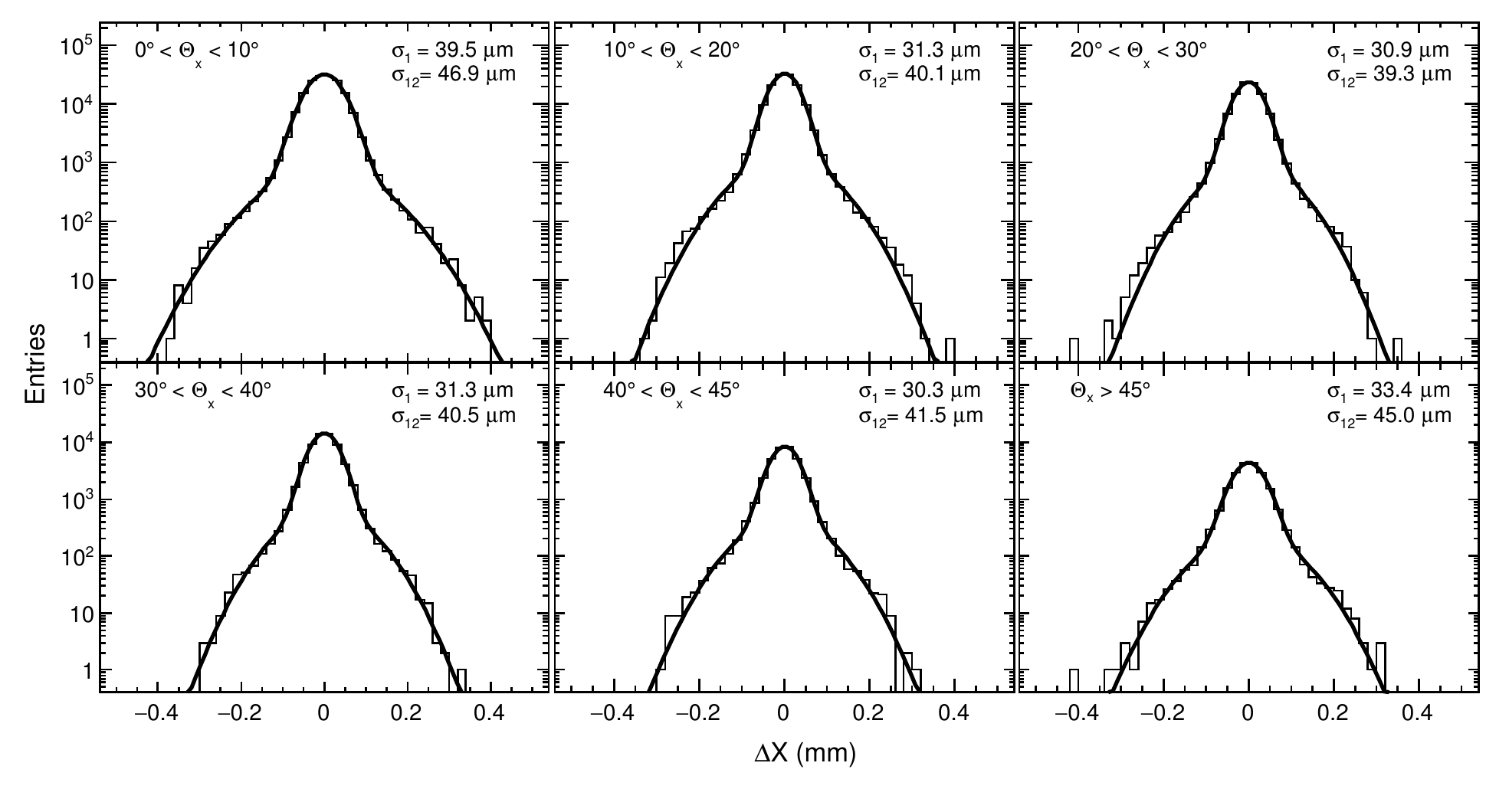}
\caption{The track-hit residual distributions for \MIP ~candidates for internal $x$ layers of the STK, shown for different track inclinations. Histograms for layers 2--5 are aggregated and fitted with the double-Gaussian distribution.}  
\label{fig:residue_x_layers_1}
\end{figure}

\begin{figure}[]
\includegraphics[width=1.0\textwidth]{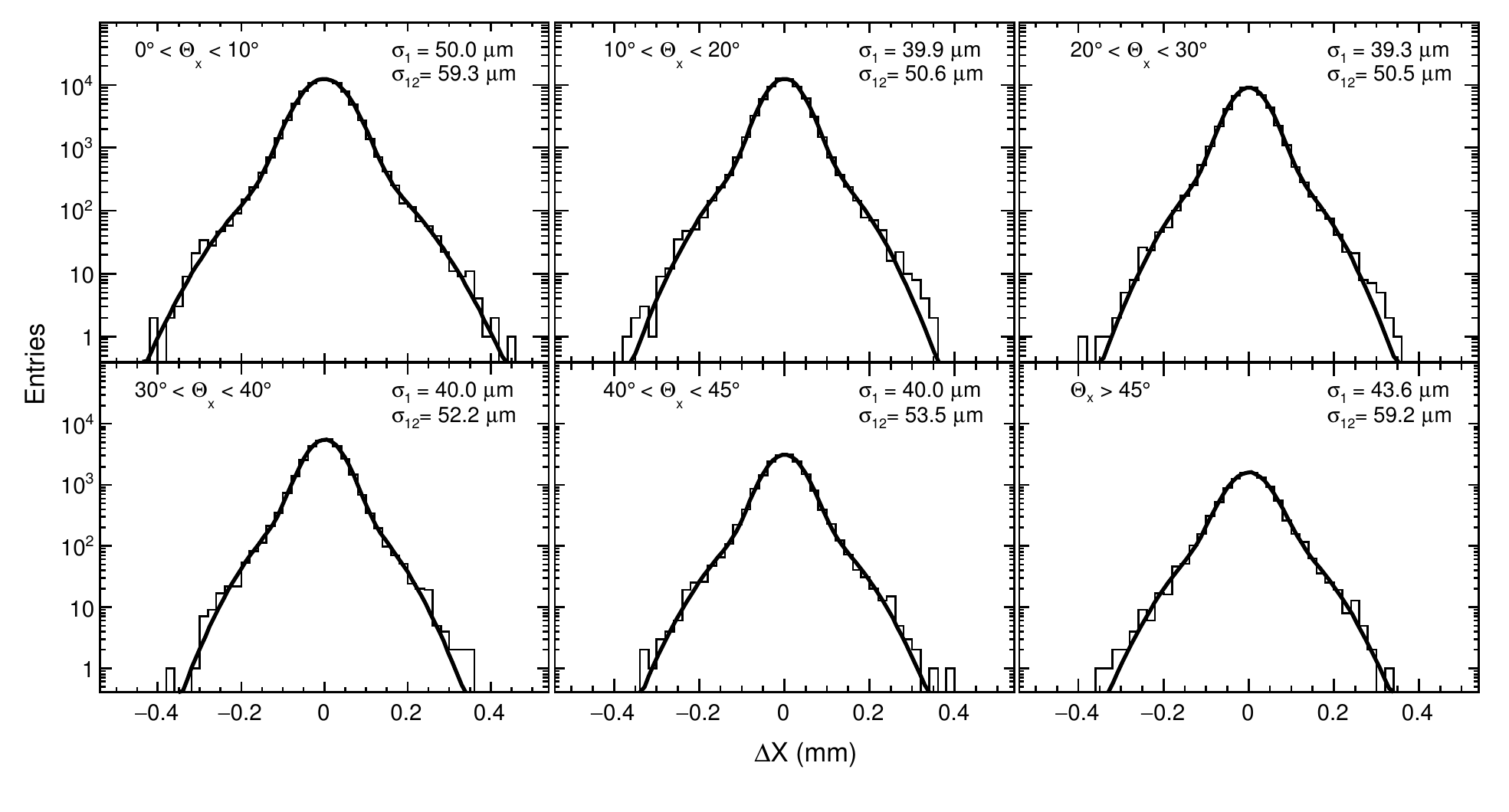}
\caption{The track-hit residual distributions for \MIP ~candidates for external $x$ layers of the STK, shown for different track inclinations. Histograms for layers 1 and 6 are aggregated and fitted with the double-Gaussian distribution.}  
\label{fig:residue_x_layers_0}
\end{figure}

The resulting track-hit residual distributions are shown as histograms of different angular ranges in Figures~\ref{fig:residue_x_layers_1} and ~\ref{fig:residue_x_layers_0} for \emph{internal} (2--5) and \emph{external} (1, 6) $x$ layers of the STK. The residual distributions for the $y$ layers show similar behavior. 
 The external layers are treated separately, since, as expected, they exhibit larger residuals due to an increased contribution of the track-projection errors for these layers. 
The histograms correspond to the data of three months, where alignment constants were updated  every two weeks, as described in Section~\ref{sec:stability}.

\begin{figure}[]
\begin{center}
\includegraphics[width=0.85\textwidth]{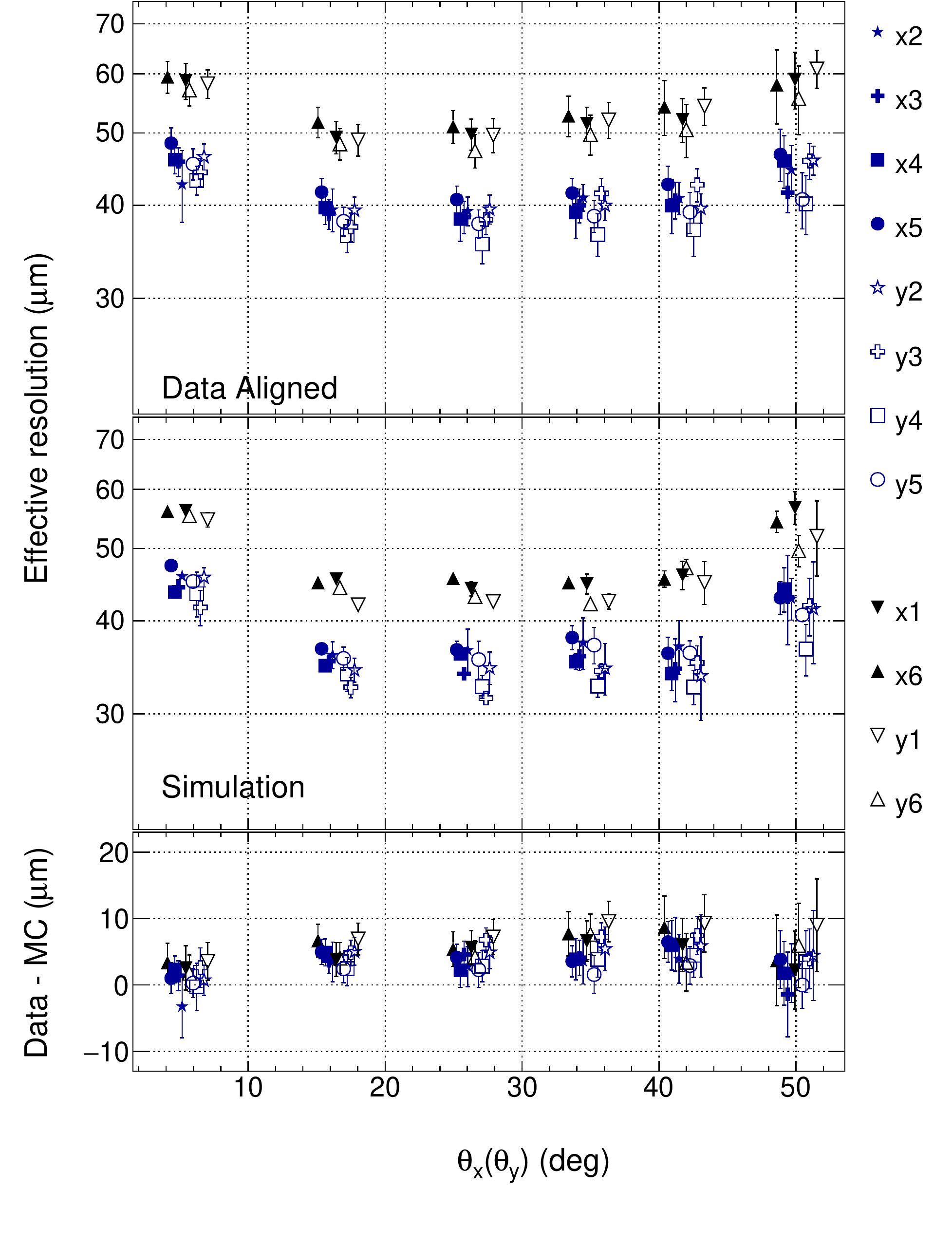}
\caption{The effective position resolution for different $x$ and $y$ planes of the STK estimated with proton candidates, and defined as the RMS ($\sigma_{12}$) of the double-Gaussian fit of the track-hit residual distributions.~\errorbarsdescribtion.}
\label{fig:pos_resolution}
\end{center}
\end{figure}

The residual distributions can be fitted with a sum of two Gaussians: 
\begin{equation}
N(x^{fit}-x^{hit}) = \frac{N_1}{\sqrt{2\pi}\sigma_1}e^{-\frac{(x^{fit}-x^{hit})^2}{2\sigma_1^2}} + \frac{N_2}{\sqrt{2\pi}\sigma_2}e^{-\frac{(x^{fit}-x^{hit})^2}{2\sigma_2^2}}
\label{eq:two_gaus}
\end{equation}
\begin{equation}
\sigma_{12} =\sqrt{\frac{N_1\sigma_1^2+N_1\sigma_2^2}{N_1+N_2}}
\end{equation}
where  $\sigma_1$ and $\sigma_2$ indicate the widths of the narrower and wider Gaussian respectively, and $\sigma_{12}$ indicates the area-weighted average width of the two Gaussians. As already described in~\cite{Alpat:2010zz}, the double-Gaussian shape of the residual distributions is due to the dependence of the position resolution on the relative position of the particle in the gap between two consecutive readout strips. The first Gaussian has a normalization ($N_1$) on average about seven times higher than the second one ($N_2$). The values of $\sigma_1$ and $\sigma_{12}$ parameters of the double-Gaussian fits in Figures~\ref{fig:residue_x_layers_1} and~\ref{fig:residue_x_layers_0} are reported in Table~\ref{tab:pos_res_p}.  Systematic uncertainties of these parameters are described in Section~\ref{sec:syst}. Figure~\ref{fig:pos_resolution} shows the effective position resolution estimated as a $\sigma_{12}$ of the double Gaussian fit for all $x$ and $y$-layers of the STK. The increase of the position resolution at low angles can be explained by the fact that particles at vertical incidence angles experience less signal sharing between readout strips, compared to inclined particles, and therefore have higher uncertainty in their position determination. A good agreement is observed between the effective position resolution obtained after the alignment and the one obtained for the ideal geometry in the simulation.  
  The apparent discrepancies of up to \mbox{10~\micron} at high incidence angles (Figure~\ref{fig:pos_resolution}) can be accounted for by uncertainties related to the charge sharing. The latter exhibits some residual differences when comparing data with simulated events.  
  Finally, Table~\ref{tab:residuals_before_after} summarizes the mean and RMS values of the residual distributions before and after the alignment for each STK layer. 

\begin{table}
\begin{center}
\begin{tabular}{cc@{$\pm$}c@{$\pm$}c@{$\pm$}cc@{$\pm$}c@{$\pm$}c@{$\pm$}c}
\hline
$\theta_x$ & ~~~~~$\sigma_1$~ & ~$\Delta_{\mathrm{stat}}$~ & ~$\Delta_{\mathrm{orb}}$~ & ~$\Delta_{\mathrm{MS}}$~  & ~~~~~$\sigma_{12}$~ & ~$\Delta_{\mathrm{stat}}$~ & ~$\Delta_{\mathrm{orb}}$~ & ~$\Delta_{\mathrm{MS}}$~   \\
\hline
\multicolumn{7}{c}{Layers 2--5}\\
\hline
             $<$10$^{\circ}$ & 39.5 &  0.8 &  1.6 &  0.4 & 46.9 &  0.7 &  1.9 &  0.8 \\
10$^{\circ}$ -- 20$^{\circ}$ & 31.3 &  0.7 &  1.3 &  0.5 & 40.1 &  0.3 &  1.6 &  0.9 \\
20$^{\circ}$ -- 30$^{\circ}$ & 30.9 &  0.7 &  1.2 &  0.6 & 39.3 &  0.2 &  1.6 &  1.1 \\
30$^{\circ}$ -- 40$^{\circ}$ & 31.3 &  0.9 &  1.3 &  0.7 & 40.5 &  0.4 &  1.6 &  1.5 \\
40$^{\circ}$ -- 45$^{\circ}$ & 30.3 &  1.2 &  1.3 &  0.7 & 41.5 &  0.8 &  1.7 &  1.9 \\
             $>$45$^{\circ}$ & 33.4 &  1.7 &  1.3 &  0.7 & 45.0 &  1.6 &  1.8 &  2.5 \\
 
\hline
\multicolumn{7}{c}{Layers 1,6}\\
\hline
             $<$10$^{\circ}$ & 50.0 &  0.1 &  2.0 &  0.8 & 59.3 &  0.1 &  2.4 &  1.6 \\
10$^{\circ}$ -- 20$^{\circ}$ & 39.9 &  0.1 &  1.6 &  1.2 & 50.6 &  0.4 &  2.0 &  1.8 \\
20$^{\circ}$ -- 30$^{\circ}$ & 39.3 &  0.1 &  1.6 &  1.5 & 50.5 &  0.6 &  2.0 &  2.2 \\
30$^{\circ}$ -- 40$^{\circ}$ & 40.0 &  0.1 &  1.6 &  1.9 & 52.2 &  0.7 &  2.1 &  2.8 \\
40$^{\circ}$ -- 45$^{\circ}$ & 40.0 &  0.1 &  1.6 &  2.3 & 53.5 &  0.6 &  2.1 &  3.6 \\
             $>$45$^{\circ}$ & 43.6 &  0.2 &  1.7 &  2.7 & 59.2 &  0.4 &  2.4 &  4.5 \\

\hline
\end{tabular}
\caption{The $\sigma_1$ and $\sigma_{12}$ of the double-Gaussian fit of the track-hit residual distributions for different $x$ planes of the STK, estimated with proton candidates. The corresponding statistical uncertainties are quoted, together with systematic uncertainties due to on-orbit variation of the alignment and the multiple scattering.~\highzuncertainty.}
\label{tab:pos_res_p}
\end{center}
\end{table}

\begin{table}
\begin{center}
\begin{tabular}{c r r r r}
\hline
STK plane & \multicolumn{2}{c}{Residuals before alignment} & \multicolumn{2}{c}{Residuals after alignment} \\
          &  Mean (\micron) & RMS (\micron)       &  Mean (\micron) & RMS (\micron)  \\
\hline
$x_1$ &  137.6~~ &  96.5~~ & 0.1~~ &  55.0~~ \\
$x_2$ &  -100.2~~ & 164.6~~ & 0.0~~ &  44.1~~ \\
$x_3$ &  -33.3~~ &  52.8~~ & -0.1~~ &  41.8~~ \\
$x_4$ &   15.6~~ & 171.0~~ & 0.0~~ &  41.7~~ \\
$x_5$ &   17.3~~ & 167.7~~ & -0.1~~ &  42.5~~ \\
$x_6$ &   16.4~~ & 332.8~~ & 0.2~~ &  53.6~~ \\
$y_1$ &   91.1~~ & 433.9~~ & 0.0~~ &  51.7~~ \\
$y_2$ &  -39.4~~ &  56.9~~ & 0.1~~ &  41.0~~ \\
$y_3$ &   25.3~~ & 197.3~~ & 0.0~~ &  38.9~~ \\
$y_4$ &  -97.2~~ & 365.7~~ & -0.1~~ &  41.2~~ \\
$y_5$ &  -13.4~~ & 108.5~~ & -0.1~~ &  42.1~~ \\
$y_6$ &  110.5~~ & 412.1~~ & 0.2~~ &  53.5~~ \\
\hline
\end{tabular}
\end{center}
\caption{The mean and RMS values of the track-hit residual distributions for proton candidates for different STK planes. The residual distributions include tracks in the whole range of incidence angles.}
\label{tab:residuals_before_after}
\end{table}
 
 % 7 times higher, on average -- to be more precise (comment from Andrii)

\subsection{Position resolution with helium}
\label{sec:resolution_helium}

The event selection is the same as the one used for protons, with the exception of the requirement on the cluster charge. 
 In order to select helium candidates, the signal of each hit is required to be in the range from 120 to ~\mbox{450~ADC} counts (see Figure~\ref{fig:p_he_selection} for the cluster charge distribution). If at least one hit in the track does not pass this selection, the track is discarded.
 This selection has 78\% efficiency for helium, with a residual proton contamination less than 0.05\%, estimated from the Monte-Carlo simulation. The energy cut in the BGO is also modified,~$E>100$~GeV, to limit the effect of higher (compared to protons) multiple scattering.  About \mbox{0.07~$\mathrm{M}$} events per day in the data pass this selection.
 
\begin{figure}[]
\includegraphics[width=1.0\textwidth]{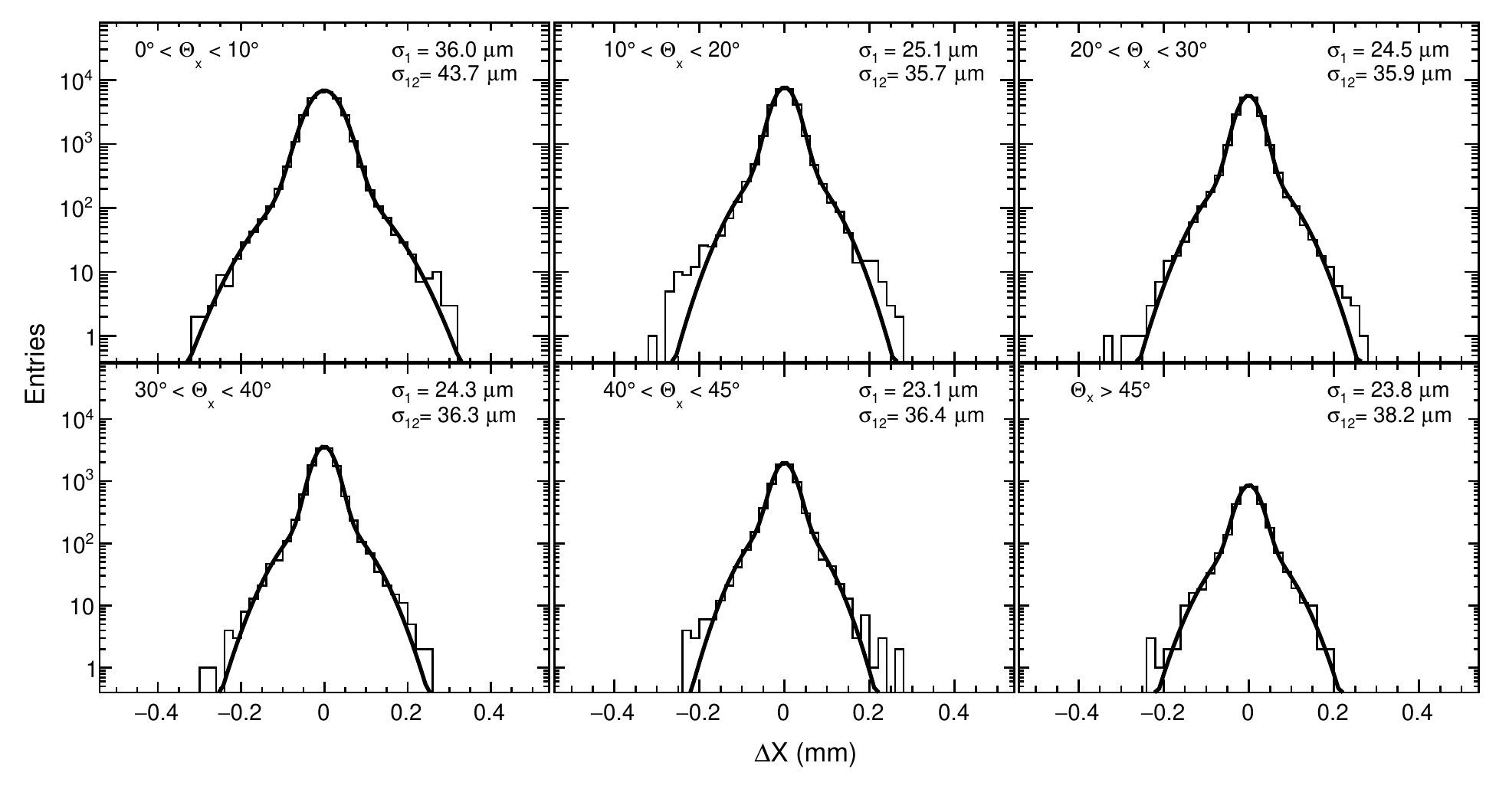}
\caption{The track-hit residual distributions for helium ion candidates for internal $x$ layers of the STK, shown for different track inclinations. Histograms for layers 2--5 are aggregated and fitted with the double-Gaussian distribution.}  
\label{fig:residue_x_helium_layers_1}
\end{figure}

\begin{figure}[]
\includegraphics[width=1.0\textwidth]{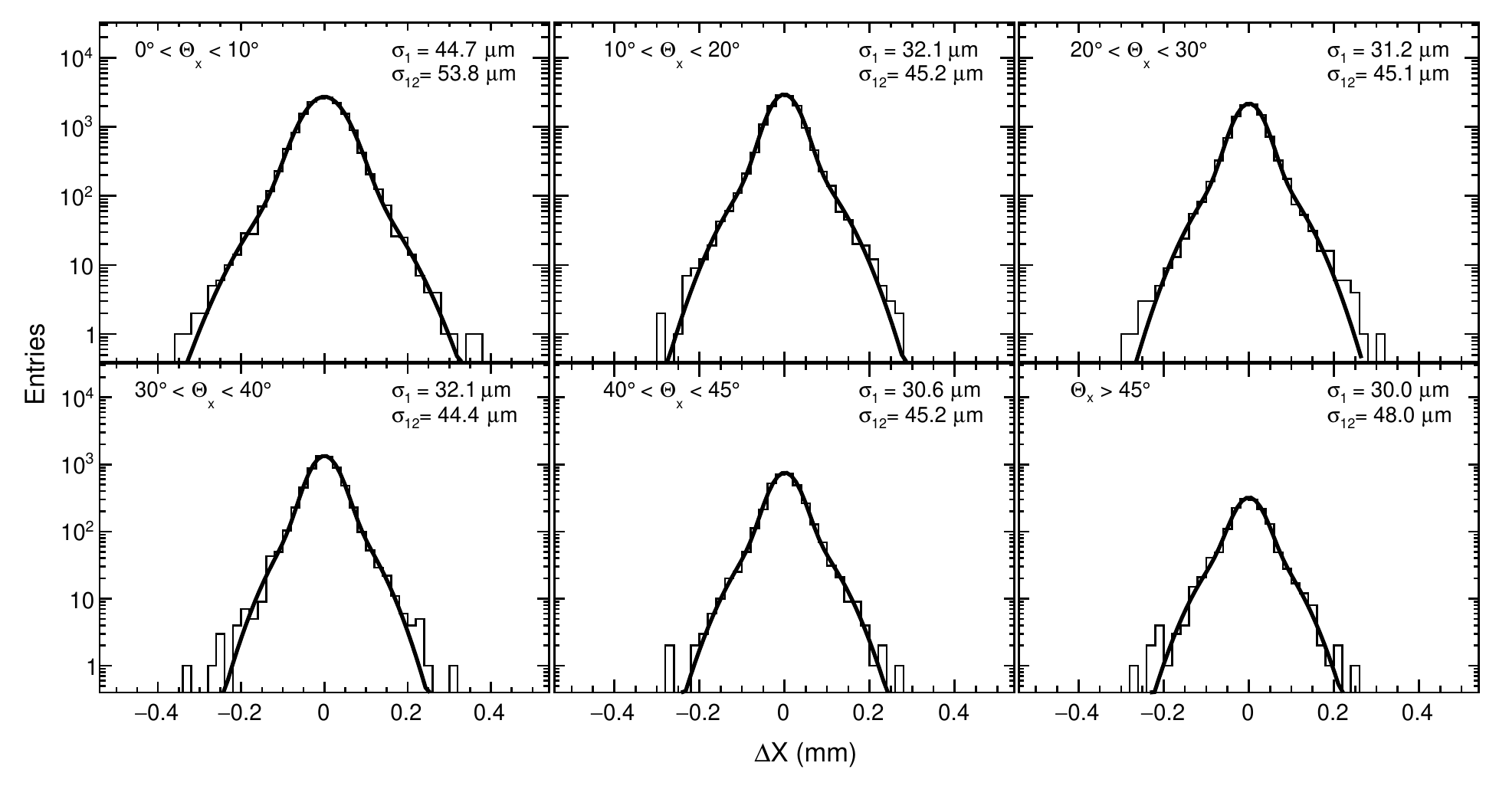}
\caption{The track-hit residual distributions for helium ion candidates for external $x$ layers of the STK, shown for different track inclinations. Histograms for layers 1 and 6 are aggregated and fitted with the double-Gaussian distribution.}  
\label{fig:residue_x_helium_layers_0}
\end{figure}

\begin{figure}[]
\begin{center}
\includegraphics[width=0.85\textwidth]{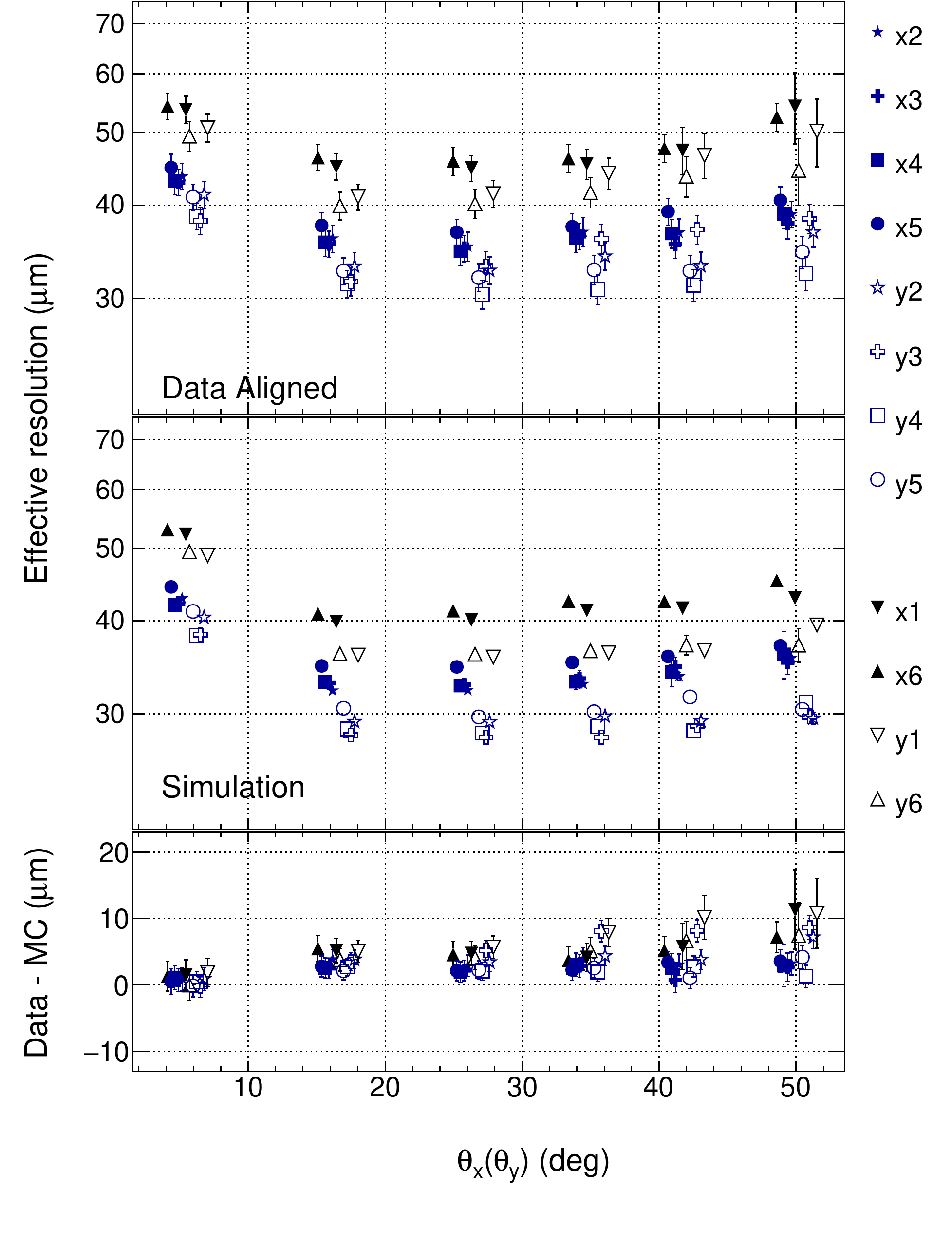}
\caption{The effective position for different $x$ and $y$ planes of the STK estimated with helium candidates, and defined as the RMS ($\sigma_{12}$) of the double-Gaussian fit of the track-hit residual distributions.~\errorbarsdescribtion. }
\label{fig:pos_resolution_helium}
\end{center}
\end{figure}

  The resulting track-hit residual distributions for helium candidates are shown in Figures~\ref{fig:residue_x_helium_layers_1} and ~\ref{fig:residue_x_helium_layers_0} for internal (2--5) and external (1,6) $x$ layers of the STK respectively, and associate fit values for the $\sigma_1$ and $\sigma_{12}$ parameters of the double-Gaussian functions are reported in Table~\ref{tab:pos_res_he}. The residuals for the $y$ layers exhibit a similar behavior.   The position resolution  for all STK layers is shown in Figure~\ref{fig:pos_resolution_helium}. It is on average up to~\mbox{5 \micron}~better than the corresponding position resolution for protons,  because of the higher (on average 4 times) signal yield.

\begin{table}
\begin{center}
\begin{tabular}{cc@{$\pm$}c@{$\pm$}c@{$\pm$}cc@{$\pm$}c@{$\pm$}c@{$\pm$}c}
\hline
$\theta_x$ & ~~~~~$\sigma_1$~ & ~$\Delta_{\mathrm{stat}}$~ & ~$\Delta_{\mathrm{orb}}$~ & ~$\Delta_{\mathrm{MS}}$~  & ~~~~~$\sigma_{12}$~ & ~$\Delta_{\mathrm{stat}}$~ & ~$\Delta_{\mathrm{orb}}$~ & ~$\Delta_{\mathrm{MS}}$~   \\
\hline
\multicolumn{7}{c}{Layers 2--5}\\
\hline
 
             $<$10$^{\circ}$ & 36.0 &  0.2 &  1.4 &  0.2 & 43.7 &  0.7 &  1.7 &  0.3 \\
10$^{\circ}$ -- 20$^{\circ}$ & 25.1 &  0.2 &  1.0 &  0.1 & 35.7 &  0.4 &  1.4 &  0.2 \\
20$^{\circ}$ -- 30$^{\circ}$ & 24.5 &  0.3 &  1.0 &  0.2 & 35.9 &  0.2 &  1.4 &  0.3 \\
30$^{\circ}$ -- 40$^{\circ}$ & 24.3 &  0.5 &  1.0 &  0.3 & 36.3 &  0.2 &  1.5 &  0.6 \\
40$^{\circ}$ -- 45$^{\circ}$ & 23.1 &  0.6 &  0.9 &  0.5 & 36.4 &  0.4 &  1.5 &  1.0 \\
             $>$45$^{\circ}$ & 23.8 &  0.8 &  1.0 &  0.8 & 38.2 &  0.8 &  1.5 &  1.6 \\
\hline
\multicolumn{7}{c}{Layers 1,6}\\
\hline
 
               $<$10$^{\circ}$ & 44.7 &  0.1 &  1.8 &  0.5 & 53.8 &  0.9 &  2.2 &  0.4 \\
10$^{\circ}$ -- 20$^{\circ}$ & 32.1 &  0.1 &  1.3 &  0.3 & 45.2 &  0.5 &  1.8 &  0.4 \\
20$^{\circ}$ -- 30$^{\circ}$ & 31.2 &  0.2 &  1.3 &  0.5 & 45.1 &  0.4 &  1.8 &  0.8 \\
30$^{\circ}$ -- 40$^{\circ}$ & 32.1 &  0.3 &  1.3 &  1.1 & 44.4 &  0.6 &  1.8 &  1.6 \\
40$^{\circ}$ -- 45$^{\circ}$ & 30.6 &  0.5 &  1.2 &  2.1 & 45.2 &  1.0 &  1.8 &  2.7 \\
             $>$45$^{\circ}$ & 30.0 &  0.7 &  1.2 &  3.6 & 48.0 &  1.7 &  1.9 &  4.1 \\
\hline
\end{tabular}
\caption{The $\sigma_1$ and $\sigma_{12}$ of the double-Gaussian fit of the track-hit residual distributions  for different $x$ planes of the STK, estimated with helium ion candidates. The corresponding statistical uncertainties are quoted, together with systematic uncertainties due to on-orbit variation of the alignment and the multiple scattering.~\highzuncertainty.}
\label{tab:pos_res_he}
\end{center}
\end{table}

\subsection{Systematic uncertainties}
\label{sec:syst}

The following sources of systematic uncertainties were considered in the analysis of the spatial resolution:
\begin{itemize}
\item Systematic uncertainty due to the contribution of multiple scattering to the residuals. It was estimated by reducing the energy selection requirement for the event selection in Sections~\ref{sec:resolution_protons} and ~\ref{sec:resolution_helium} by a factor of two, down to 25 GeV for protons and 50 GeV for helium nuclei. The resulting difference in position resolution with respect to the baseline selection reaches at most 2.5 (1.6)~\micron~ for protons (helium nuclei) for internal layers of the STK, and 4.5 (4.1)~\micron~ for protons (helium nuclei) for the external layers.
\item Systematic uncertainty due to alignment variation on orbit. It was estimated as a maximum variation with time of the RMS of narrower Gaussian ($\sigma_1$) in the fit of a track-hit residual distribution, as described in Section~\ref{sec:stability}. %It is conservatively 
 This uncertainty is conservatively taken to be 
 within 4\%, which corresponds to a maximum uncertainty of up to 2.4~\micron~and 2.2~\micron~for protons and helium nuclei respectively.
\item Systematic uncertainty due to the contribution of nuclei with charge $Z>1$ in the alignment sample, estimated by adding the proton charge requirement to the alignment event selection, ~\hitsignal~$<$~160~ADC counts.  A helium contamination in this case was found below 0.1\%, estimated from the Monte-Carlo simulation by fitting the proton and helium charge distribution to the data, as shown in Figure~\ref{fig:p_he_selection}.
 The position resolution estimation was repeated and compared with the baseline, and the resulting difference was taken as a systematic uncertainty. It is less than 0.1~\micron~and 0.2~\micron~for $\sigma_1$ and $\sigma_{12}$ respectively.
\end{itemize}

Finally, the statistical uncertainties of the data and Monte-Carlo samples were estimated by splitting each sample in two statistically independent parts and performing the double-Gaussian fit for each part separately. The resulting difference divided by $\sqrt{2}$ was taken as a statistical uncertainty. It varies depending on a track incidence angle and reaches at most 2~\micron~for both proton and helium data. For the proton and helium Monte-Carlo it was found to be less than~8~\micron~and 4~\micron~respecrively.

\section{Alignment stability}
\label{sec:stability}

To assess the stability of the alignment,  we use the \emph{control} data sample  obtained with the STK-based event selection, as follows.  The residuals of a linear fit of 5 points in the projection to the 6-\emph{th} plane (point under study) are required to be lower than ~\mbox{40 \micron}.  Residuals are then evaluated with a similar methodology to the one described in Section~\ref{sec:resolution}, and fitted with a double-Gaussian distribution. The \emph{control} selection helps  to reduce the contribution of multiple scattering in the track-hit residuals, while allowing to collect sufficient statistics (about~\mbox{2$~\mathrm{M}$} tracks) in less than 5 days of data. 
 To avoid any bias due to the chosen selection criteria, we have also checked the same selection with a  lower threshold of~\mbox{10 \micron}. The resulting variation of the estimated position resolution is below 10$\%$.  We also use the \emph{control} event selection to validate the convergence of the alignment algorithm and to estimate the required statistics of data for the alignment. As a measure of the stability of the alignment, we use the width of the narrower Gaussian of a fit of track-hit residual distribution, $\sigma_1$ (see Equation~\ref{eq:two_gaus}).  

\subsection{Alignment stability with time}
\begin{figure}[]
\includegraphics[width=1.0\textwidth]{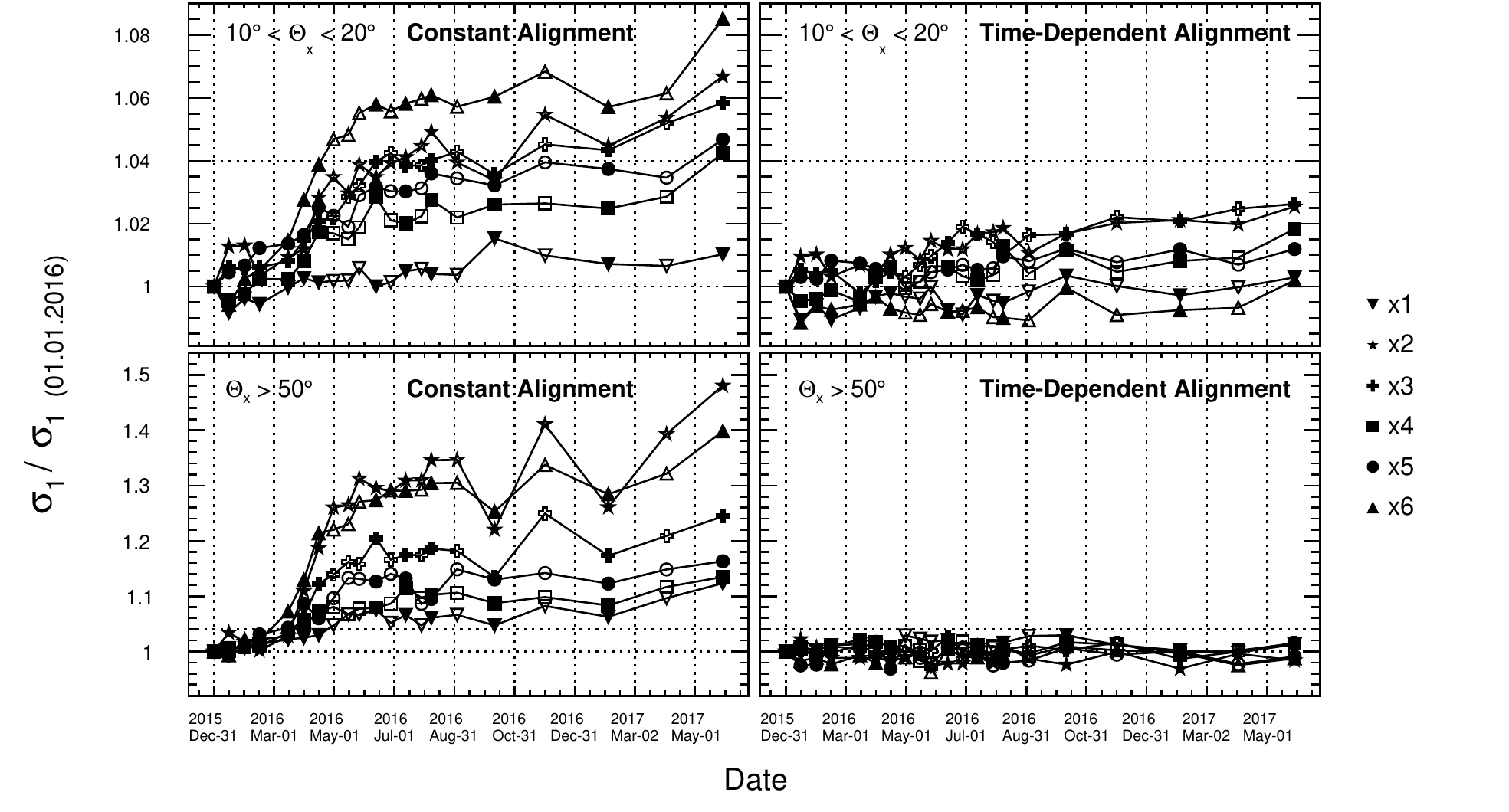}  % Stephan
\caption{The variation of the $\sigma_1$ for different STK planes at different track inclinations as a function of time. %The $\sigma_1$ values are normalized to their corresponding values at January 1$^{\mathrm{st}}$ 2016. 
 The values are shown for $x$ planes only, while  $y$ planes show similar behavior. Horizontal dashed lines are shown to indicate 0 and 4\% deviation from the initial values. Two cases are shown, the \emph{constant} alignment (left) and the \emph{time-dependent} alignment (right). In the first case, a single alignment is performed on January 1$^{\mathrm{st}}$ 2016. In the second one, the alignment is performed at least once every two weeks of the data taking. A sparse subset of points for dates after August 2016 is shown  for legibility. }
\label{fig:alignment_stability}
\end{figure}

 Here we examine the position resolution for different time periods.  In Figure~\ref{fig:alignment_stability} we show the variation of the $\sigma_1$  parameter as a function of time for both the fixed and the time-dependent alignment methods. In the first case, the alignment parameters have been computed only for January 2016. In the second one, those parameters are updated once  every two weeks and the  closest ones in time are used for each event in the~\emph{control} sample.  As shown in Figure~\ref{fig:alignment_stability}, in case of the time-dependent alignment, the variation of the position resolution with time is less than 4\%.  We have also checked the effect of performing the alignment once every two days and once every week, without finding any significant improvement compared to the baseline procedure.

\begin{figure}[]
\begin{centering}
\includegraphics[width=1.0\textwidth]{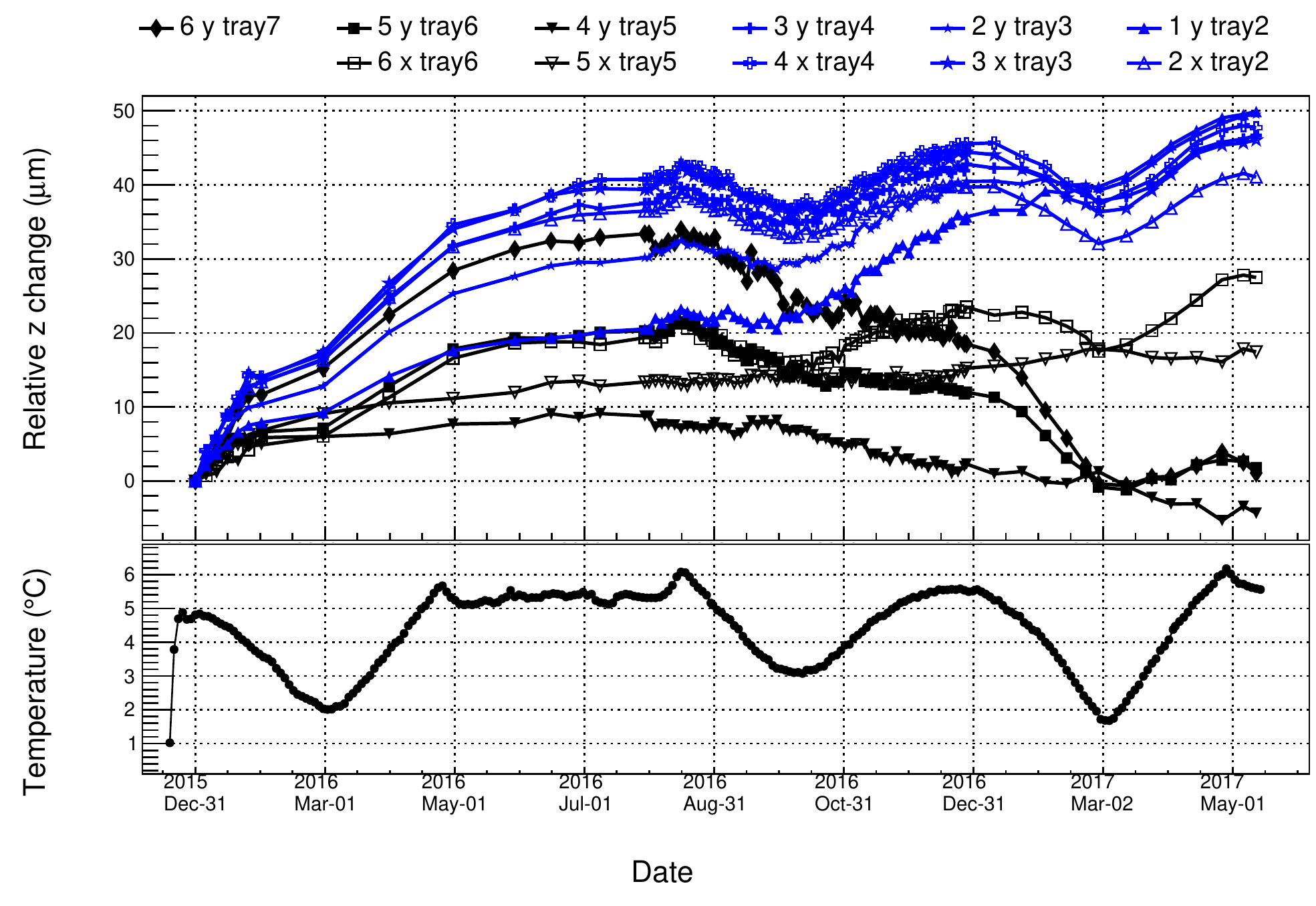}
\caption{The variation of the average $z$ position of each STK tracking layer with respect to the first layer (6$x$, tray 1) as a function of time (top), and the average temperature of the STK, measured by 384 temperature sensors installed on the STK ladders (bottom). During the period between July 2016 and December 2016 realignment was done every three days, in order to examine the alignment variation with higher timing resolution. The blue curves correspond to the trays with the tungsten plates.}
\label{fig:alignment_variation}
\end{centering}
\end{figure}

\begin{table}
\begin{center}
\begin{tabular}{c r c c}
\hline
%STK mechanical part  & Dimension &Temperature dependence coefficient  & Moisture dependence coefficient \\
STK part & Dimension (mm) & $k_T$ ~$\left(\frac{\mathrm{\mu m}}{^\circ C}\right)$  & $k_H$~$\left(\frac{\mathrm{\mu m}}{\%\mathrm{H_2 O}}\right)$\\
\hline
\multicolumn{4}{c}{In-plane direction}\\
\hline
Ladder flex & 380~~~~~  & 7.6  &  0.4\\
Corner feet & 181~~~~~ &  4.2 & 0.0\\
Support tray & 940~~~~~ & 0.9 & 9.4\\
SSD & 95~~~~~  & 0.3 & 0.0\\
\hline
\multicolumn{4}{c}{Out-of-plane direction}\\
\hline
Corner feet  & 236~~~~~ & 5.5 & 0.0\\
Support tray & 28~~~~~ & 0.0 & 0.3\\
\hline
\end{tabular}
\end{center}
\caption{Conservative limits on the temperature and humidity dependence coefficients, $k_{T}$ and $k_{H}$, for different mechanical components of the STK:   the aluminum corner feet, the support trays composed of carbon fiber and aluminum honeycomb, %mounted on the corner feet,  
 PCB ladder flexes and the silicon sensors. The temperature and humidity expansion can be parameterized as $L=L_0 + k_{T}(T-T_{0}) + k_{H}(H-H_{0})$, where $L$ is a linear dimension of an STK part, $T$ is temperature and $H$ is humidity.}
\label{tab:temp_humd_coef}
\end{table}

Figure~\ref{fig:alignment_variation} shows the variation of average alignment parameters in the off-plane direction ($z$-axis). The change of those with time can be explained by two main effects: humidity release and temperature variation. In the beginning of on-orbit operation, the humidity release process is expected to cause  a contraction of the carbon fiber trays.  Then, the STK temperature varies up  to 4\tempdegree~due to seasonal variation of the DAMPE orbit.  As seen from Figure~\ref{fig:alignment_variation} (bottom), the maximum rate of temperature change on orbit reaches about 0.5\tempdegree~per week. 
The temperature is measured by 384 sensors installed on the STK ladders, 2 sensors per ladder. It is expected that the temperatures of the other parts of  the STK follow a similar behavior with time.
 Table~\ref{tab:temp_humd_coef} summarizes the temperature and humidity expansion coefficients for different parts of the STK. The tracking planes are mounted on the four aluminum corner feet (see Figure~\ref{fig:dampe}), which expand or contract in the $z$ direction because of the temperature variations. Moreover, temperature and humidity  expansion/contraction of trays can result in bending of trays and therefore contribute to the off-plane position change as well. 
  The $z$ position variation impacts mostly inclined tracks, since tracks with low inclination are not sensitive to the $z$ coordinate. It can be seen from the left plots in the Figure~\ref{fig:alignment_stability} that the variation of $\sigma_1$ for inclined tracks is much higher than for vertical ones.

The variation of the average alignment in $x$ and $y$ direction was found to be within 1~\micron.  This can be explained by the fact that the mechanical structure of the STK is resting on the four aluminum corner feet and four aluminum frames housing the data acquisition boards (one frame attached to each side of the STK) which prevent the trays from relative shifts in $x$ and $y$ direction.

\subsection{Performance of the alignment algorithm}

\begin{figure}[]
\begin{center}
\includegraphics[width=0.7\textwidth]{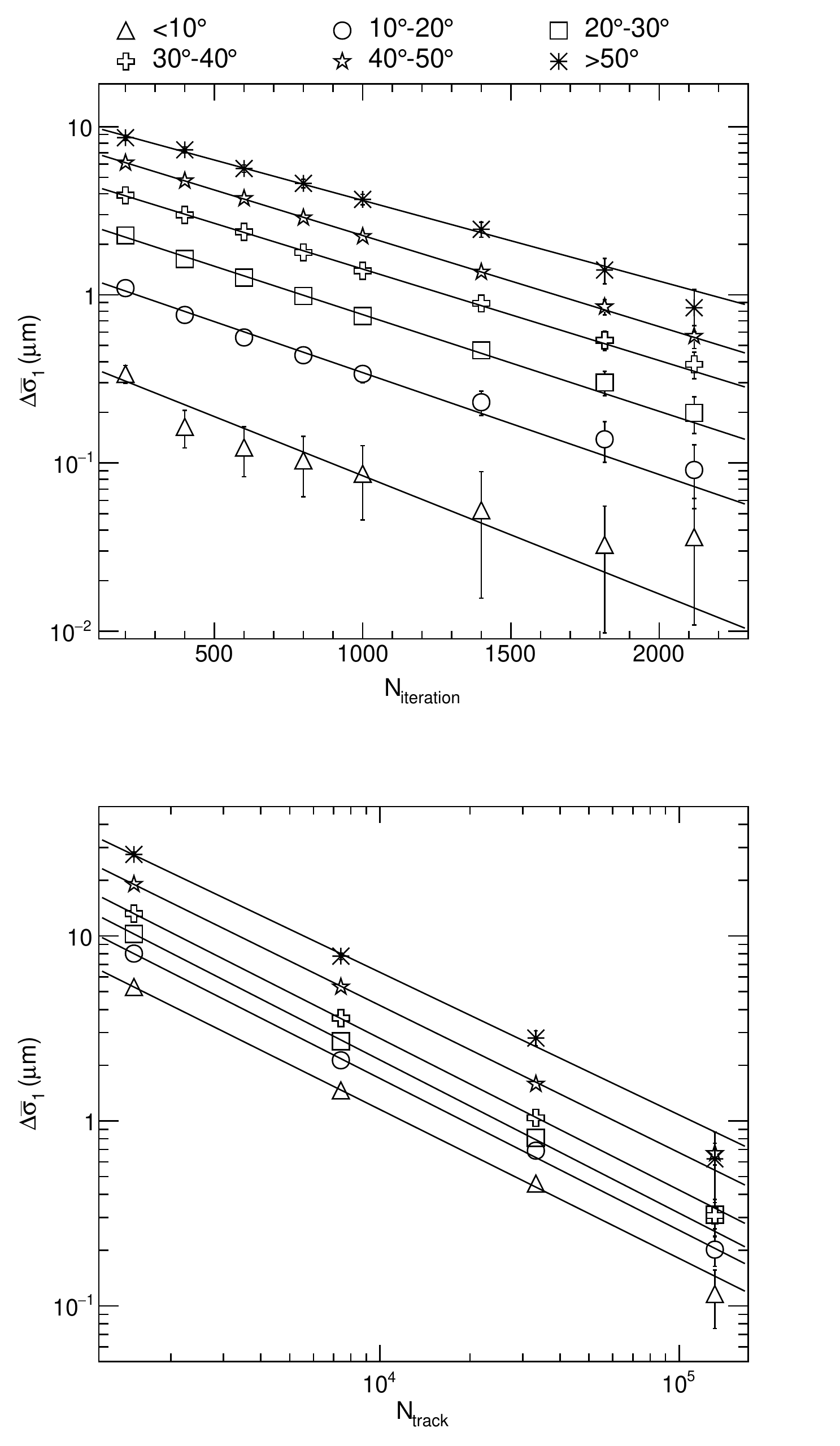} 
\caption{The variation of the average $\sigma_1$  as a function of iteration number of the alignment algorithm (top) and number of tracks used for the alignment (bottom). The average is taken for all $x$ and $y$ layers of the STK and shown for different intervals of the  particle incidence angle. The values in the top and the bottom plots are fitted with the formulae~\ref{eq:conv_iter}  and~\ref{eq:conv_ntracks} respectively.}
\label{fig:convergance_rate}
\end{center}
\end{figure}

In order to study the performance of the alignment algorithm, we evaluate the alignment parameters using either statistically reduced data samples or samples of a normal size but with a reduced number of iterations for the alignment procedure. 
 Then, we estimate the position resolution using these parameters and compare it with the one obtained with the baseline alignment procedure.   
  Figure~\ref{fig:convergance_rate}  shows the variation of the $\sigma_1$ parameter  as a function of the number of iterations for the alignment procedure and the number of tracks in the alignment sample.   The dependencies in Figure~\ref{fig:convergance_rate}  can be fitted with the following functions:
\begin{align}
\Delta\bar{\sigma}_1 &= a_\mathrm{I}  \cdot e^{-b_\mathrm{I} \cdot N_{\mathrm{iteration}}} \label{eq:conv_iter} \\
\Delta\bar{\sigma}_1 &= a_\mathrm{N} \cdot  \left( N_{\mathrm{track}} \right)  ^{-b_\mathrm{N}} \label{eq:conv_ntracks}
\end{align}
where $N_{\mathrm{iteration}}$ and $N_{\mathrm{track}}$ are the number of iterations and number of tracks used for the alignment respectively. Table~\ref{tab:convergence_fit_results} summarizes the fit values for these functions. As seen from the table, the values of the convergence slope $b_\mathrm{N}$ for different incidence angles ($\mathrm{\theta_X}$/$\mathrm{\theta_Y}$) are compatible with one another within the statistical uncertainties. On the other hand, the convergence slope $b_\mathrm{I}$ decreases slowly as a function of the $\mathrm{\theta_X}$/$\mathrm{\theta_Y}$. It can be explained by the fact that the statistics of tracks in the alignment sample decreases with the growth of $\mathrm{\theta_X}$/$\mathrm{\theta_Y}$. Therefore, the tracks with low incidence angle contribute more to the total $\chi^2$ (Equation~\ref{eq:chisq_track_2}).  

\begin{figure}[]
\begin{center}
\includegraphics[width=1.0\textwidth]{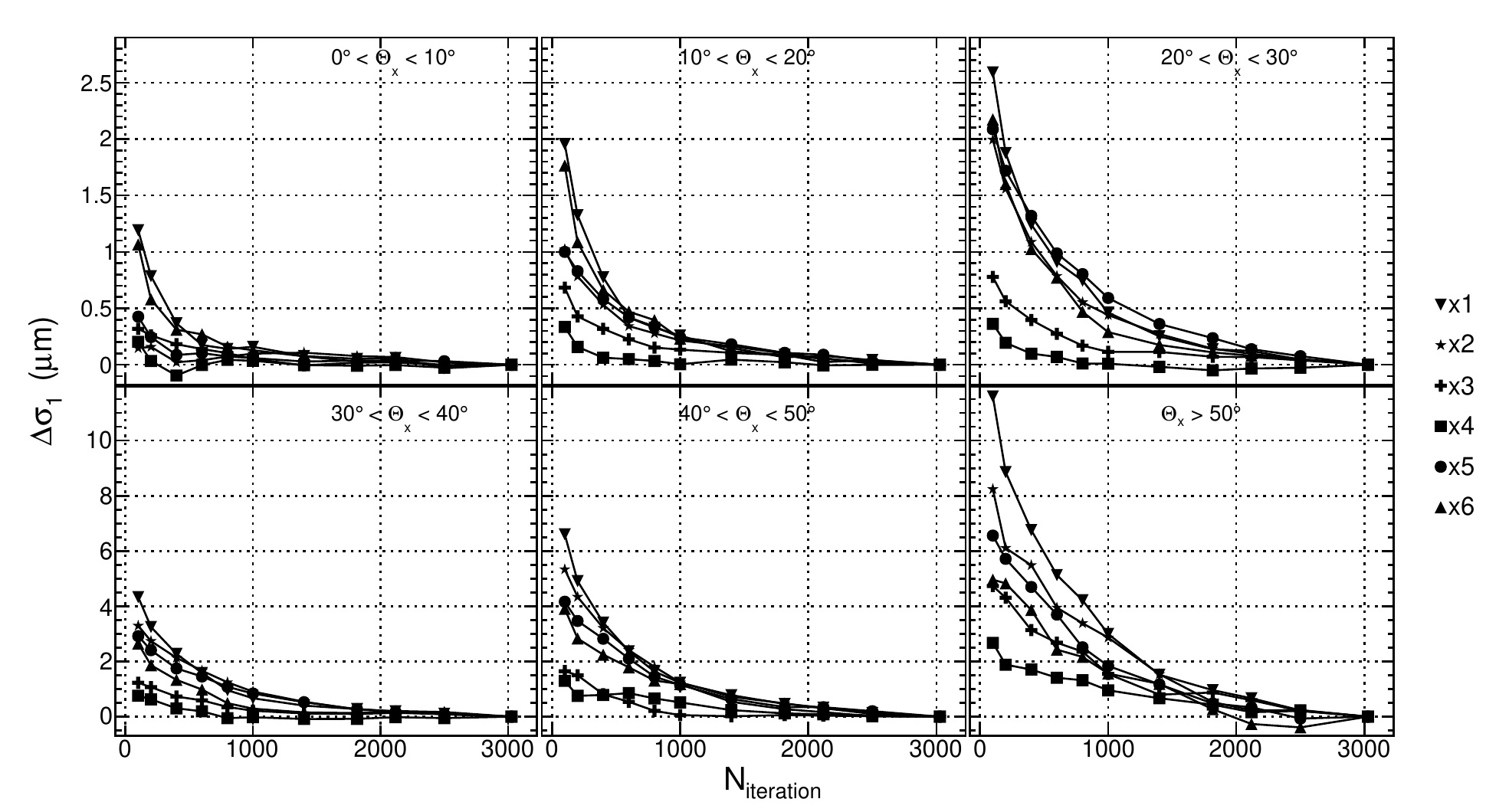} % Stephan
\includegraphics[width=1.0\textwidth]{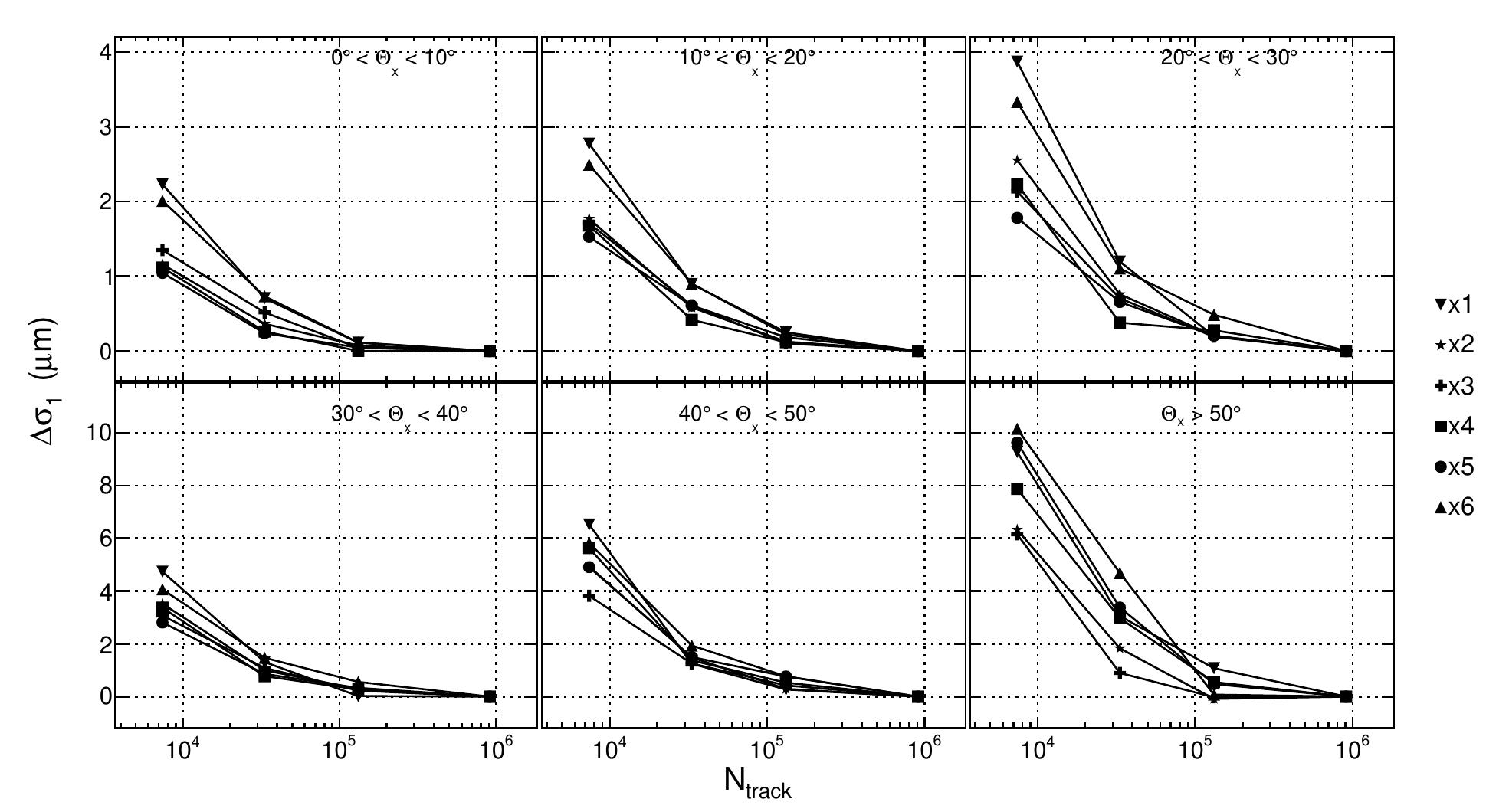}
\caption{The variation of the $\sigma_1$ for each $x$ layer of STK as a function of iteration number of the alignment algorithm (top) and number of tracks used for the alignment (bottom).}
\label{fig:convergance}
\end{center}
\end{figure}

\begin{table}
\begin{center}
\begin{tabular}{c r@{~$\pm$~}l r@{~$\pm$~}l r@{~$\pm$~}l r@{~$\pm$~}l}
\hline
$\mathrm{\theta_X}$/$\mathrm{\theta_Y}$  & \multicolumn{4}{c}{ Convergence for $N_{\mathrm{iteration}}$ }  & \multicolumn{4}{c}{Convergence for $N_{\mathrm{track}}$} \\
&  \multicolumn{2}{c}{$a_\mathrm{I}$ (\micron)} & \multicolumn{2}{c}{$b_\mathrm{I}$ $\times$ 10$^{-3}$} &   \multicolumn{2}{c}{$a_\mathrm{N}$ (\micron) $\times$ 10$^{3}$} & \multicolumn{2}{c}{$b_\mathrm{N}$ $\times$ 10$^{-1}$} \\
\hline
               $<$10$^{\circ}$ &   0.42 &    0.07 &   1.62 &   0.30 &   1.92 &    0.26 &   8.05 &   0.17 \\
    10$^{\circ}$--20$^{\circ}$ &   1.39 &    0.06 &   1.40 &   0.08 &   3.25 &    0.28 &   8.20 &   0.11 \\
    20$^{\circ}$--30$^{\circ}$ &   2.87 &    0.08 &   1.32 &   0.05 &   4.39 &    0.40 &   8.28 &   0.12 \\
    30$^{\circ}$--40$^{\circ}$ &   4.98 &    0.11 &   1.25 &   0.04 &   5.32 &    0.55 &   8.20 &   0.13 \\
    40$^{\circ}$--50$^{\circ}$ &   7.85 &    0.15 &   1.25 &   0.03 &   6.42 &    0.60 &   7.96 &   0.12 \\
               $>$50$^{\circ}$ &  11.02 &    0.41 &   1.11 &   0.06 &   7.71 &    1.67 &   7.71 &   0.24 \\
\hline
\end{tabular}
\caption{Parameters of the Equations~\ref{eq:conv_iter} and~\ref{eq:conv_ntracks}, obtained from the fits in Figure~\ref{fig:convergance_rate}.}
\label{tab:convergence_fit_results}
\end{center}
\end{table}

As seen from Figures~\ref{fig:convergance_rate}~top and~\ref{fig:convergance}~top, around 2000 iterations are needed to obtain the best alignment, especially for the highly inclined tracks,  which are particularly sensitive to the alignment in the off-plane ($z$-axis) direction. It should be noted, however, that less than 100 iterations are needed for the in-flight realignment, since the alignment procedure is applied to the already pre-aligned geometry. As a convergence criterion, we require the variation of average $\sigma_1$ to be below~\mbox{1~\micron}.   Finally, as shown in Figures~\ref{fig:convergance_rate}~bottom and~\ref{fig:convergance}~bottom,  about~\mbox{1~$\mathrm{M}$} tracks is sufficient to perform the alignment, which roughly corresponds to 1--2 days of data.

\iffalse
\begin{figure}[]
\begin{center}
\includegraphics[width=0.75\textwidth]{figure14.pdf} % Stephan
\caption{Variation of the $\sigma_1$ for each $x$ layer of STK as a function of the iteration number of the alignment algorithm.}
\label{fig:convergance}
\end{center}
\end{figure}

\begin{figure}[]
\begin{center}
\includegraphics[width=0.75\textwidth]{figure15.pdf}
\caption{Variation of $\sigma_1$ for each $x$ layer of STK as a function of number of tracks used for the alignment.}
\label{fig:statistics}
\end{center}
\end{figure}
\fi

%\clearpage
\section{Conclusions}
\label{sec:conclusion}

During its first 17 months of on-orbit operation, DAMPE has collected 2.6 billion events. Using this data we performed a detailed study and optimization of the alignment procedure of the silicon  tracker. Given the mechanical stability of the tracker structure and the limited temperature variation of the detector, the alignment parameters are updated every two weeks to ensure optimal tracking performance.
We estimate the effective position resolution for protons in the $x$ and $y$ internal tracker planes, after the alignment procedure is applied, to be \mbox{47$\pm$2~\micron} for events arriving at normal incidence ($<$10$^{\circ}$), \mbox{41$\pm$2~\micron} for intermediate inclinations and \mbox{45$\pm$3~\micron} at high incidence angles ($>$45$^{\circ}$). For helium nuclei the estimated position resolution is \mbox{44$\pm$2~\micron}, \mbox{36$\pm$2~\micron} and \mbox{38$\pm$3~\micron} at low, intermediate and high incidence angles respectively. 
 The effective position resolution in the external tracker layers is on average 12~\micron~and 9~\micron~worse than in the internal layers for protons and helium nuclei respectively, due to the high contribution of projection errors, as also observed in the simulation.
We show that the variation of the optimal position resolution with time is less than 4\%,  i.e. less than \mbox{2~\micron},  given the implemented procedure.
Results for the position resolution of the aligned tracker agree well with the results of the simulation for the nominal DAMPE geometry.

\section*{Acknowledgment}

%The DAMPE mission is funded by the strategic priority science and technology projects in space science of Chinese Academy of Sciences. In Europe  the experiment is supported by the Swiss National Science Foundation (SNSF), Switzerland and the National Institute for Nuclear Physics (INFN), Italy. The authors wish to express their gratitude to the generosity of CERN for providing beam time allocation and technical assistance at the PS and SPS beam lines, as well as general logistical support is acknowledged. 

The DAMPE mission is funded by the strategic priority science and technology projects in space science of Chinese Academy of Sciences (No.XDA04040000 and No. XDA04040400). In Europe the experiment is supported by the Swiss National Science Foundation under grant No. 200020\_175806  and the National Institute for Nuclear Physics (INFN), Italy. The authors wish to express their gratitude to the generosity of CERN for providing beam time allocation and technical assistance at the PS and SPS beam lines, as well as general logistical support is acknowledged.

%\section*{References}

\bibliography{mybibfile}

\end{document}